\documentclass[aps,prl,reprint,superscriptaddress,amsmath,amssymb]{revtex4-2}
\usepackage{graphicx}
\usepackage[colorlinks=true,linkcolor=blue,citecolor=blue,urlcolor=blue]{hyperref}
\usepackage{comment}
\newcommand{\dd}{\mathrm{d}}
\newcommand{\ee}{\mathrm{e}}
\newcommand{\Tr}{\operatorname{Tr}}

\begin{document}

\title{Quantization Breaks Classical Mpemba Constraints}

\author{Yue Liu}
\affiliation{Center for Gravitational Physics and Quantum Information, Yukawa Institute for Theoretical Physics, Kyoto University, Kitashirakawa Oiwakecho, Sakyo-ku, Kyoto 606-8502, Japan}
\author{Tan Van Vu}
\affiliation{Center for Gravitational Physics and Quantum Information, Yukawa Institute for Theoretical Physics, Kyoto University, Kitashirakawa Oiwakecho, Sakyo-ku, Kyoto 606-8502, Japan}
\author{Fr\'ed\'eric van Wijland}
\affiliation{Laboratoire Mati\`ere et Syst\`emes Complexes (MSC), Universit\'e Paris Cit\'e \& CNRS (UMR 7057), 75013 Paris, France}
\affiliation{Center for Gravitational Physics and Quantum Information, Yukawa Institute for Theoretical Physics, Kyoto University, Kitashirakawa Oiwakecho, Sakyo-ku, Kyoto 606-8502, Japan}
\author{Hisao Hayakawa}
\affiliation{Center for Gravitational Physics and Quantum Information, Yukawa Institute for Theoretical Physics, Kyoto University, Kitashirakawa Oiwakecho, Sakyo-ku, Kyoto 606-8502, Japan}

\begin{abstract}
The Mpemba effect, in which an initially hotter system relaxes faster than a colder one, has been widely studied in both classical and quantum systems, yet identifying the specific effects of quantization requires a consistent comparison with the corresponding classical dynamics. 
Starting from the Caldeira-Leggett model, we establish a quantum-classical correspondence for thermal relaxation.
We find that the locality of classical energy diffusion forbids the Mpemba effect in single-well and symmetric double-well potentials.
Conversely, quantum dynamics allows for direct transitions between nonadjacent energy levels, overcoming the constraints of classical local energy diffusion and enabling the Mpemba effect in these potentials.
In the semiclassical limit $\hbar\to0$, the nonlocal transitions become increasingly suppressed, restoring the classical constraints.
Our findings reveal that quantization fundamentally unlocks anomalous relaxation pathways, rather than merely modifying relaxation rates.
\end{abstract}
\maketitle

\textit{Introduction.---}More than two millennia ago, Aristotle noted that prior heating could accelerate the freezing of water~\cite{Aristotle1923}. 
This puzzling phenomenon was brought into modern scientific focus through the work of Mpemba and Osborne in 1969~\cite{mpemba1969cool}. 
The Mpemba effect describes the counterintuitive phenomenon in which an initially hotter system relaxes to thermal equilibrium faster than an initially colder one when both are coupled to the same cold bath~\cite{lu2017nonequilibrium,Ares2025,TEZA2026}. 
Controlled experiments across diverse physical platforms have demonstrated the Mpemba effect~\cite{kumar2020exponentially,tian2025experimental} and its quantum counterparts~\cite{Joshi2024,Shapira2024}, showing that a greater initial distance from the stationary state need not imply a longer relaxation time.
Together with other anomalous relaxation phenomena, including heating--cooling asymmetry~\cite{Lapolla2020,Ibanez2024} and the Kovacs memory effect~\cite{Santos24}, this effect underscores the critical role of initial state preparation and dynamical pathways in thermal relaxation.

The Mpemba effect has been extensively investigated in both classical~\cite{lu2017nonequilibrium,Santos17,Torrente19,klich2019mpemba,Baity-Jesi19,Yang20,Santos20,Biswas20,kumar2020exponentially,Takada21a,chetrite2021metastable,Mompo20,Busiello21,walker2021anomalous,Kumar22,Biswas21,Degunther2022,Megias22b,Holtzman22,Biswas23a,Biswas23,Santos24,Malhotra2024,Biswas24,tian2025experimental,Antonov2026,takadahayakawa,liu2026mpemba,pre-us,Shapira2026Underdamped} and quantum systems~\cite{Nava19,Carollo21,Manikandan21,Ares23,Ivander23,Chatterjee_2023,Rylands2024,Joshi2024,Shapira2024,Yamashika2024,Chatterjee2024,Wang24,Nava24,Moroder2024,Liu24,Chalas24,Longhi24,Zhang2025,Yamashika2025,Strachan2025,Turkeshi2025,Bao2025,Beato2026,Bao2026,Yamashika2026,Hayakawa2026Tunneling,Melles2026}.
These studies encompass relaxation from thermal initial states~\cite{lu2017nonequilibrium,klich2019mpemba,kumar2020exponentially,chetrite2021metastable,walker2021anomalous,Kumar22,Biswas23a,Malhotra2024,tian2025experimental,takadahayakawa,liu2026mpemba,pre-us,Hayakawa2026Tunneling,Melles2026,Shapira2026Underdamped} as well as nonequilibrium preparations~\cite{Santos17,Torrente19,Biswas20,Takada21a,Mompo20,Carollo21,Biswas21,Degunther2022,Ares23,Chatterjee_2023,Rylands2024,Joshi2024,Shapira2024,Wang24,Nava24,Moroder2024,Longhi24,Zhang2025,Beato2026}.
Spectral analyses have linked this anomalous relaxation to the interplay between initial state preparation and the dynamical relaxation modes~\cite{lu2017nonequilibrium,klich2019mpemba,chetrite2021metastable,Carollo21,walker2021anomalous,Biswas23a,Chatterjee_2023,Beato2026,liu2026mpemba,pre-us,Hayakawa2026Tunneling,Shapira2026Underdamped}.
However, the correspondence between classical and quantum Mpemba effects, particularly the quantum--classical crossover, remains a missing piece of the puzzle. 
Recent work compared classical overdamped dynamics with quantum Gorini--Kossakowski--Sudarshan--Lindblad (GKSL) dynamics in a benchmark double-well potential~\cite{Melles2026}, but the classical limit of the latter is not the former.
A major challenge lies in connecting classical and quantum relaxations within a unified system--bath framework, matching both the potential and the bath spectral density while establishing how the quantum dynamics approaches its classical counterpart. 
This leads to the central question of what role quantum effects play in Mpemba relaxation relative to the corresponding classical dynamics.


\begin{figure}[t]
\centering
\includegraphics[width=\columnwidth]{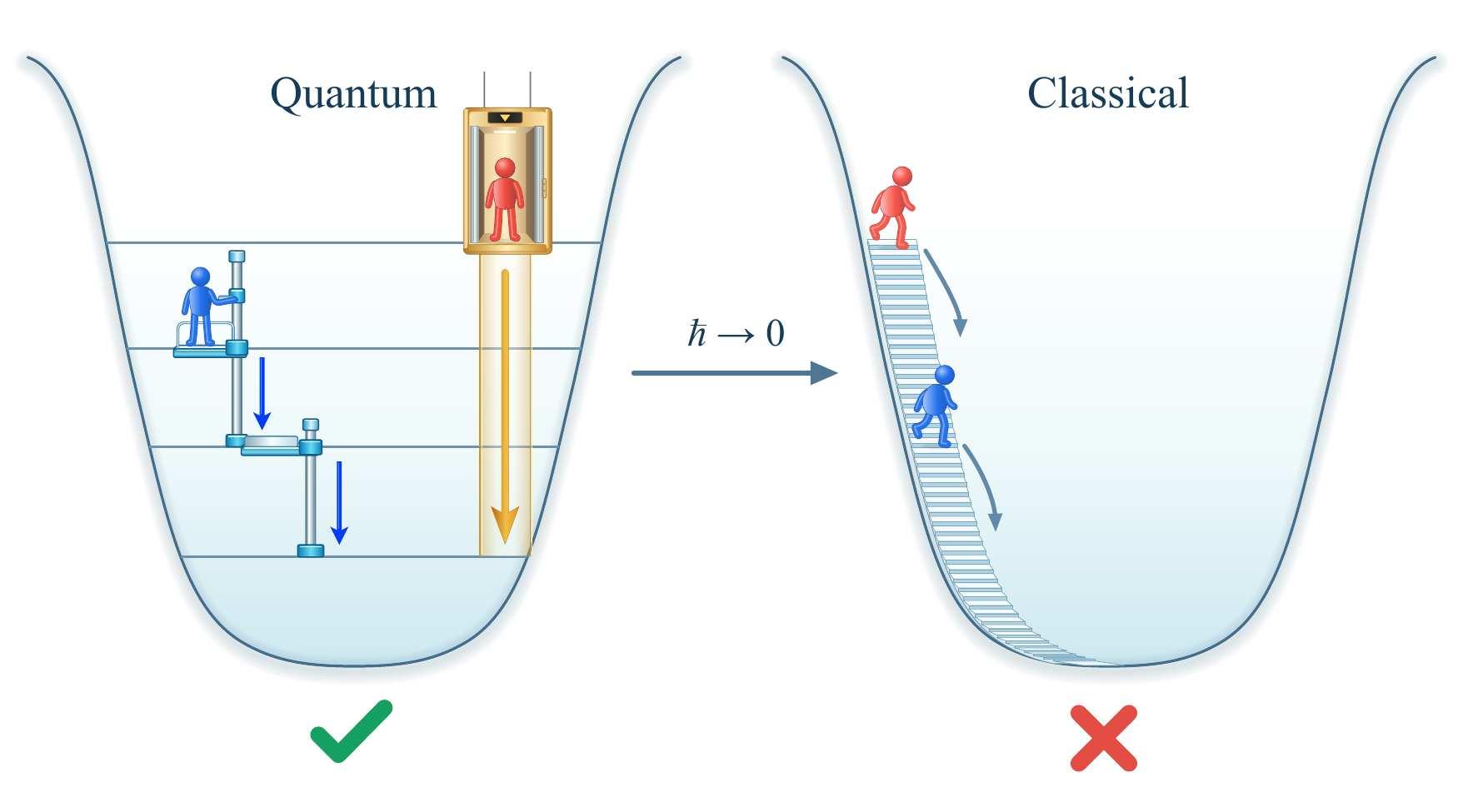}
\caption{Schematic comparison of quantum and classical relaxations.
Red and blue figures schematically represent higher and lower energy states. 
The orange elevator illustrates a nonadjacent quantum transition, while the blue moving platform indicates successive adjacent transitions.
The staircase represents local diffusion in continuous classical energy.
The arrow between the two wells indicates the classical limit.
The check and cross denote, respectively, the possible occurrence and absence of the Mpemba effect.}
\label{fig:quantum-classical-schematic}
\end{figure}

In this Letter, we address these questions using quantum and classical descriptions of thermal relaxation derived from the Caldeira--Leggett model~\cite{CaldeiraLeggett1983} in the weak-coupling regime. 
The local nature of classical energy diffusion (right panel of Fig.~\ref{fig:quantum-classical-schematic}) imposes constraints that rule out the Mpemba effect for thermal initial states in single-well and symmetric double-well potentials.
We show that transitions between nonadjacent quantum energy levels (left panel of Fig.~\ref{fig:quantum-classical-schematic}) overcome these constraints, enabling the Mpemba effect even in a symmetric single well. 
As $\hbar$ decreases, we trace quantum-classical crossovers, showing how the classical constraints are restored.

\textit{Quantum dynamics.---}We consider a particle of mass $M$ confined in a one-dimensional potential $V(x)$, with the Hamiltonian
\begin{equation}\label{eq:hamiltonian}
  \hat H=-\frac{\hbar^2}{2M}\frac{\dd^2}{\dd x^2}+V(x).
\end{equation}
The discrete energy levels $E_\mu$ and the corresponding normalized eigenstates $|\mu\rangle$ verify $\hat H|\mu\rangle=E_\mu|\mu\rangle$.
In the following, we use dimensionless quantities, with the effective Planck constant $\tilde\hbar=\hbar t_0/(M_0 L_0^2)$, where $M_0$ is a fixed mass unit (such as proton mass), $L_0$ is the characteristic length of the potential (such as the distance between the two wells).
We set $t_0$ proportional to the inverse gap between the two lowest energy levels and keep it fixed thereafter (see Supplemental Material (SM)~\cite{SM} for details).
Then the effective Planck constant can be adjusted by rescaling the potential (equivalently, varying $L_0$), which can be realized experimentally in atom-optical systems~\cite{Hensinger2001DrivenPendulum,Steck2002Fluctuations,Arnal2020ChaosAssisted}.
For brevity, we henceforth write $\hbar$ instead of $\tilde{\hbar}$.

We describe the interaction with the environment using the quantum Caldeira--Leggett model~\cite{CaldeiraLeggett1983} (see End Matter for details), in which the particle couples linearly via its position operator $\hat x$ to a bath of harmonic oscillators in thermal equilibrium at inverse temperature $\beta_b$.
In the weak-coupling regime, the Born--Markov and secular approximations yield a GKSL equation for the reduced density matrix of the system~\cite{Davies1974,Gorini1976,Lindblad1976,Breuer2002} (see SM~\cite{SM}):
\begin{equation}\label{eq:lindblad}
  \begin{aligned}
    \dot\rho={}&-\mathrm i\hbar^{-1}[\hat H,\rho] \\
    &+\sum_\omega\gamma(\omega)\left(A_\omega\rho A_\omega^\dagger-\frac12\{A_\omega^\dagger A_\omega,\rho\}\right).
  \end{aligned}
\end{equation}
The sum runs over distinct Bohr frequencies, and the corresponding jump operators are
\begin{equation}\label{eq:jump-operators}
  A_\omega=\sum_{E_\nu-E_\mu=\hbar\omega}x_{\mu\nu}|\mu\rangle\langle\nu|,\qquad x_{\mu\nu}=\langle\mu|\hat x|\nu\rangle.
\end{equation}
Crucially, these jump operators connect nonadjacent energy levels, as illustrated for the quantum case in Fig.~\ref{fig:quantum-classical-schematic}, with transition amplitudes determined by the position matrix elements $x_{\mu\nu}$.
For $\omega>0$, the dissipative coefficient is $\gamma(\omega)=J(\omega)\hbar^{-1}\left[n_{\mathrm B}(\omega)+1\right]$, where $J(\omega)$ is the bath spectral density, and $n_{\mathrm B}(\omega)=[\exp(\beta_b\hbar\omega)-1]^{-1}$ is the Bose--Einstein distribution at the bath inverse temperature $\beta_b$.
The coefficient satisfies the detailed balance condition $\gamma(-\omega)=\exp(-\beta_b\hbar\omega)\gamma(+\omega)$.

We consider an initial thermal state $\rho(0)=Z_q(\beta)^{-1}\exp(-\beta\hat H)$ at inverse temperature $\beta$, with partition function $Z_q(\beta)=\Tr \exp(-\beta\hat H)$.
This mixed state is diagonal in the energy eigenbasis, and the dynamics in Eq.~\eqref{eq:lindblad} preserves this diagonal structure throughout relaxation.
It suffices to track the populations $p_\mu(t)=\langle\mu|\rho(t)|\mu\rangle$, which obey the Pauli master equation~\cite{Breuer2002},
\begin{equation}\label{eq:pauli}
  \dot p_\mu=\sum_{\nu\ne\mu}\left[\Gamma_{\mu\leftarrow\nu}p_\nu-\Gamma_{\nu\leftarrow\mu}p_\mu\right]\equiv(\mathcal L_q\mathbf p)_\mu,
\end{equation}
where $\mathbf p(t)=(p_0(t),p_1(t),\ldots)^{\mathsf T}$ is the population vector and $\mathcal L_q$ is the quantum population Liouvillian. 
The initial populations are given by $\pi_\mu(\beta)=Z_q(\beta)^{-1}\exp(-\beta E_\mu)$, and the transition rates are $\Gamma_{\mu\leftarrow\nu}=|x_{\mu\nu}|^2\gamma[\hbar^{-1}(E_\nu-E_\mu)]$ for $\nu\ne\mu$.

\textit{Classical dynamics.---}We now consider the classical limit of the Pauli dynamics in Eq.~\eqref{eq:pauli}.
Eliminating the bath degrees of freedom from the classical Caldeira--Leggett model leads to a generalized Langevin equation~\cite{Zwanzig1973}.
In the weak-damping regime, coarse-graining over bath memory and averaging over the fast degrees of freedom yield the energy diffusion equation~\cite{KRAMERS1940,Zwanzig1959,CarmeliNitzan1983},
\begin{equation}\label{eq:classical-energy-diffusion}
  \begin{aligned}
    \partial_t P(E,t)&=\partial_E\left[D_E(E)\pi(E;\beta_b)\partial_E\left(\frac{P(E,t)}{\pi(E;\beta_b)}\right)\right]\\
    &\equiv\mathcal L_c P(E,t),
  \end{aligned}
\end{equation}
where $P(E,t)$ is the probability density in energy space, $\mathcal L_c$ is the classical Liouvillian, and $\pi(E;\beta_b)$ is the equilibrium energy density at the bath inverse temperature $\beta_b$. 
The thermal energy density is given by $\pi(E;\beta)=Z_c(\beta)^{-1}\ee^{-\beta E}g(E)$, where $Z_c(\beta)$ ensures normalization and $g(E)$ is the density of states with energy $E$.
The energy diffusion coefficient is given by~\cite{CarmeliNitzan1983,KarrleinGrabert1998}
\begin{equation}\label{eq:energy-diffusion-coefficient}
  D_E(E)=\beta_b^{-1}\sum_{k=1}^{\infty}k\Omega(E)J\!\left(k\Omega(E)\right)|x_k(E)|^2,
\end{equation}
where $\Omega(E)$ is the orbital angular frequency given by $2\pi/g(E)$ (see Eq.~(S46) in SM~\cite{SM}) and $x_k(E)$ is the $k$th complex Fourier coefficient of the position along the classical trajectory.
With energy $E$ rather than position $x$ or momentum $p$ as the slow variable, the possibility of the Mpemba effect depends on the structure of the dynamics in energy space.
Importantly, Eq.~\eqref{eq:classical-energy-diffusion} is a distinct description in the weak-damping regime, rather than an equivalent reformulation of the underdamped Kramers or overdamped Smoluchowski equations (see End Matter for details).
As illustrated on the classical side of Fig.~\ref{fig:quantum-classical-schematic}, the energy diffusion dynamics in Eq.~\eqref{eq:classical-energy-diffusion} exchanges population only between neighboring energies, in sharp contrast to the nonlocal energy exchange on the quantum side.
The dynamics in Eq.~\eqref{eq:classical-energy-diffusion} can be realized experimentally~\cite{Turlot1989,Gieseler2014,Rondin2017} in nanoscopic systems.

Several remarks are in order. 
First, the classical Caldeira--Leggett model follows from its quantum counterpart in the limit $\hbar\to0$. 
Weak damping, coarse-graining over bath memory, and orbital averaging play roles analogous to weak coupling, the Born--Markov approximation, and secular averaging on the quantum side.
Second, Eq~\eqref{eq:classical-energy-diffusion} can be independently obtained by taking the limit $\hbar\to0$ of the Pauli dynamics~\cite{KarrleinGrabert1998} (see SM~\cite{SM} for details).
Consequently, for the same mass, potential, bath temperature, and bath spectral density, the classical energy diffusion in Eq.~\eqref{eq:classical-energy-diffusion} constitutes the exact classical counterpart of the quantum Pauli dynamics in Eq.~\eqref{eq:pauli}. 
Third, depending on the approximations employed, the quantum Caldeira--Leggett model admits several reduced dynamical descriptions, each of which has a classical counterpart (see End Matter).

\textit{Mpemba effect.---}Both the quantum and classical dynamics can be written in the unified form
\begin{equation}\label{eq:general-dynamics}
  \dot p(t)=\mathcal Lp(t),
\end{equation}
where $p$ denotes either the quantum population vector $\mathbf p$ or the classical energy density $P$, and $\mathcal L$ represents the corresponding Liouvillian $\mathcal L_q$ or $\mathcal L_c$.
Similarly, $\pi(\beta)$ denotes the thermal distribution at inverse temperature $\beta$, with components $\pi_\mu(\beta)$ for the quantum case and density $\pi(E;\beta)$ for the classical case.
Here, we consider the relaxation process from thermal initial states $\pi(\beta)$ to the bath equilibrium state $\pi(\beta_b)$.

The operator $\mathcal L$ in Eq.~\eqref{eq:general-dynamics} has nonpositive eigenvalues $\{-\lambda_n\}_{n\ge1}$, with $0=\lambda_1<\lambda_2<\lambda_3<\cdots$. 
We represent the corresponding right and left eigenmodes by $r_n$ and $\ell_n$, normalized by $\langle \ell_m,r_n\rangle=\delta_{mn}$, where the pairing is a sum over quantum energy levels or an integral over classical energy.
For a thermal initial state, the solution of Eq.~\eqref{eq:general-dynamics} can be expanded as
\begin{equation}
  p(t)=\pi(\beta_b)+\sum_{n\ge2}a_n(\beta)\ee^{-\lambda_n t}r_n,
\end{equation}
where the amplitude of the $n$th relaxation mode is given by the overlap with the left eigenmodes,
\begin{equation}\label{eq:thermal-mode-amplitudes}
  a_n(\beta)=\langle \ell_n,\pi(\beta)\rangle,
\end{equation}
which quantifies how strongly the initial state excites that relaxation mode.

At long times, relaxation is dominated by the slowest relaxation mode, and the deviation from equilibrium then takes the form $p(t)-\pi(\beta_b)\simeq a_2(\beta)\ee^{-\lambda_2t}r_2$. 
The decay rate and relaxation mode are fixed by the dynamics, leaving $a_2(\beta)$ to determine the dependence on the initial state. 
A smaller $|a_2(\beta)|$ therefore implies a smaller late-time deviation from equilibrium for standard measures of distance~\cite{Tan2025}, such as the Kullback--Leibler (KL) divergence and the $L^1$ distance.
For two thermal initial states with $\beta_h<\beta_c<\beta_b$, the hotter state is initially farther from equilibrium but becomes closer to it at sufficiently long times if $|a_2(\beta_h)|<|a_2(\beta_c)|$. 
An analogous reversal can occur during heating, with the initially colder state approaching equilibrium faster~\cite{lu2017nonequilibrium}.
Such reversals define the Mpemba effects, characterized by a nonmonotonic dependence of $a_2(\beta)$.
A necessary condition for this nonmonotonicity is that the derivative
\begin{equation}\label{eq:thermal-overlap-derivative}
  \frac{\dd a_2(\beta)}{\dd\beta}=-\operatorname{Cov}_{\pi(\beta)}(E,\ell_2) 
\end{equation}
vanishes and changes sign as the initial temperature varies. 
Here, the covariance is $\operatorname{Cov}_{\pi(\beta)}(E,\ell_2)=\langle E\ell_2\rangle_{\pi(\beta)} -\langle E\rangle_{\pi(\beta)}\langle\ell_2\rangle_{\pi(\beta)}$, with $\langle\cdot\rangle_{\pi(\beta)}=\langle\cdot,\pi(\beta)\rangle$, and the energy variable $E$ denotes $E_\mu$ in the quantum case and the continuous energy in the classical case.



\textit{Classical constraints.---}Investigating the four classes of confining potentials reveals a striking contrast, as summarized in Table~\ref{tab:quantum-classical-mpemba-possibility}.
The Mpemba effect can occur in all four classes under quantum Pauli dynamics, whereas classical energy diffusion permits it only in asymmetric double wells. 
A check mark indicates that the effect is possible for a suitable choice of potential, system parameters, and bath spectrum, while a cross denotes its absence for any potential in that class and any choice of system parameters or bath spectrum within the regime of validity of Eq.~\eqref{eq:classical-energy-diffusion}.
Quantum dynamics therefore enables the effect in three classes of potentials where its classical counterpart forbids it.

\begin{table}[t]
\centering
\begin{tabular}{lc@{\;}c@{\;}c}
  \hline
  Potential class & Quantum & $\xrightarrow{\hbar\to0}$ & Classical \\
  \hline
  Symmetric single well   & $\checkmark$ & & $\times$ \\
  Asymmetric single well  & $\checkmark$ & & $\times$ \\
  Symmetric double well   & $\checkmark$ & & $\times$ \\
  Asymmetric double well  & $\checkmark$ & & $\checkmark$ \\
  \hline
\end{tabular}
\caption{Possibility of the Mpemba effect under the quantum dynamics in Eq.~\eqref{eq:pauli} and the corresponding classical dynamics in Eq.~\eqref{eq:classical-energy-diffusion}.}
\label{tab:quantum-classical-mpemba-possibility}
\end{table}

We first establish the classical constraint for single-well potentials.
In this case, the relaxation modes satisfy the Sturm--Liouville eigenvalue problem~\cite{risken1996fokker}, which implies that $\ell_2(E)$ has exactly one node.
Multiplying its eigenvalue equation by $\pi(\beta_b)$ and integrating from
the lower boundary gives
\begin{equation}\label{eq:classical-slow-mode-monotonicity}
  D_E(E)\pi(E;\beta_b)\ell_2'(E)=-\lambda_2\int_{E_{\min}}^E\pi(u;\beta_b)\ell_2(u)\,\dd u,
\end{equation}
where $E_{\min}$ is the minimum energy of the well.
The single node and the orthogonality condition $\langle\ell_2,\pi(\beta_b)\rangle=0$ ensure that the integral in Eq.~\eqref{eq:classical-slow-mode-monotonicity} has a fixed, nonzero sign for all $E>E_{\min}$, which means that $\ell_{2}(E)$ is strictly monotonic.
The one-dimensional Fortuin--Kasteleyn--Ginibre inequality~\cite{fortuin1971correlation} guarantees that $\operatorname{Cov}_{\pi(\beta)}(E,\ell_2)$ is strictly sign-definite and nonvanishing as $\beta$ varies.
By Eq.~\eqref{eq:thermal-overlap-derivative}, $\dd a_2/\dd\beta$ can neither vanish nor change sign, forbidding the Mpemba effect in both symmetric and asymmetric single-well potentials.
Kramers energy diffusion, a special case of Eq.~\eqref{eq:classical-energy-diffusion} for an Ohmic bath in the white-noise limit (see Fig.~\ref{fig:dynamical-framework} in End Matter for details), also excludes the Mpemba effect in single-well potentials~\cite{Shapira2026Underdamped}.
This contrasts with memoryless overdamped dynamics, where confining walls enable the Mpemba effect in single-well potentials~\cite{liu2026mpemba,pre-us}.
For the classical energy diffusion dynamics considered here, a single-well potential excludes the effect, regardless of the presence of walls.
These results show that different descriptions of classical Brownian motion can yield qualitatively different Mpemba behavior even in the same potential.

For a symmetric double-well potential, symmetry keeps the two wells equally populated throughout relaxation from a thermal initial state.
The symmetric dynamics can thus be mapped onto one half of the potential with a reflecting boundary at the symmetry point~\cite{pre-us}, allowing the single-well result to apply directly and exclude the Mpemba effect.
For an asymmetric double-well potential, the energy space has distinct branches below the barrier and we lose the constraint that $\ell_2$ varies monotonically with energy (see Fig.~(S3) in SM~\cite{SM}).
States above the barrier  contribute less to slow relaxation than metastable states in the shallower well. 
A higher initial temperature can then place more population in these higher states, reducing $a_2(\beta)$ and allowing for the Mpemba effect.
Because we focus on the quantum-classical correspondence here, the numerical illustrations of the nonmonotonic behavior of $a_2(\beta)$ and a detailed discussion of the asymmetric double-well potential are sent to SM~\cite{SM}.

\begin{figure}[t]
\centering
\includegraphics[width=\columnwidth]{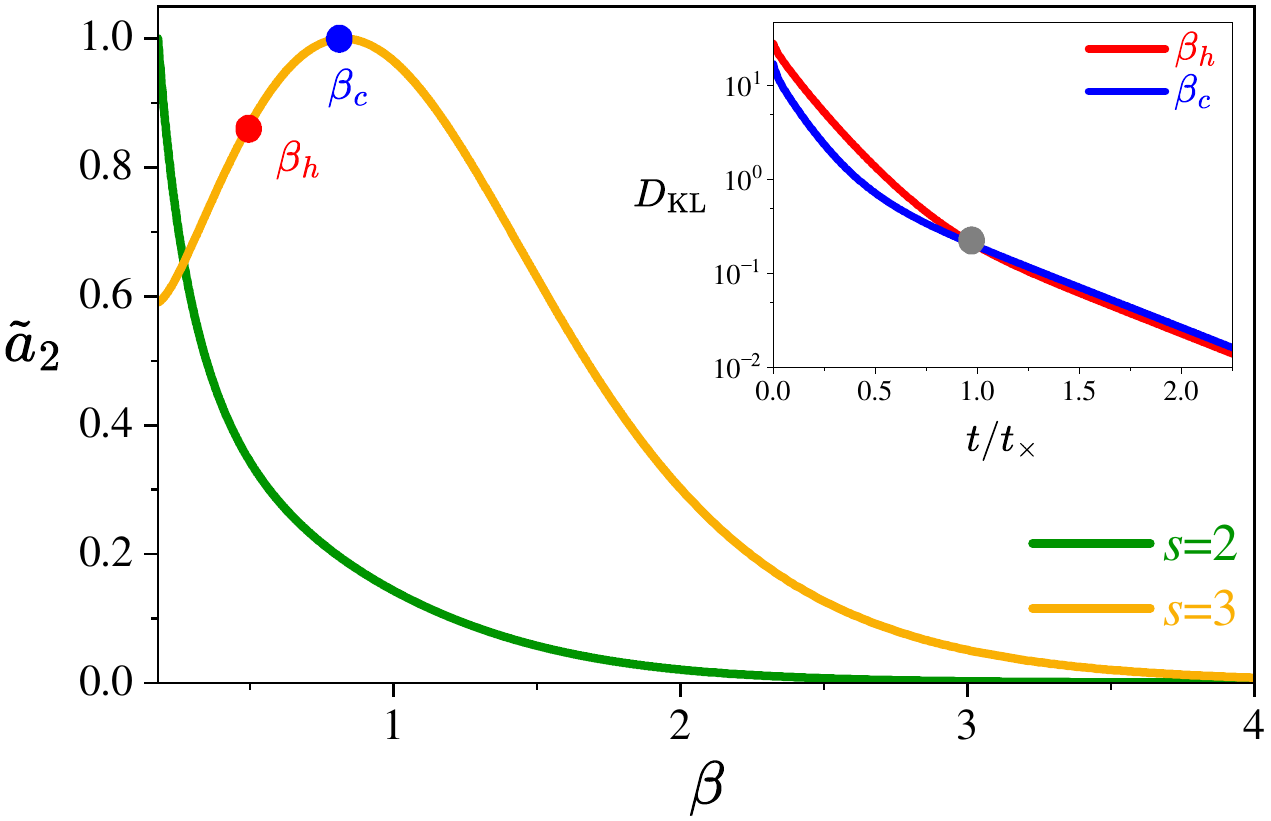}
\caption{Normalized overlap coefficient $\tilde a_2=a_2/(\max_\beta |a_2|)$ as a function of the initial inverse temperature $\beta$ for $s=2$ and $s=3$, with each maximum taken over the plotted interval.
Blue and red markers indicate $\beta_c=5/6$ and $\beta_h=1/2$, respectively.
The inset shows the corresponding KL relaxation curves for the indicated initial states with $s=3$, where time is scaled by their crossing time $t_\times$.
Here, $\hbar^2/(2M)=0.14$, $\hbar=1$, $\eta=1$, $\beta_b=10$ and $\omega_c=8$.}
\label{fig:quantum-mpemba}
\end{figure}

The classical constraints result from local diffusion, where population is exchanged only between neighboring energies. 
As illustrated in the classical case in Fig.~\ref{fig:quantum-classical-schematic}, population moving from high to low energies must pass through the intermediate energies, just as a person goes down a staircase step by step.

\textit{Quantum Mpemba effect.---}In the quantum case, energy takes discrete values and relaxation proceeds through transitions between energy levels.
According to Eq.~\eqref{eq:jump-operators}, the jump operators connect nonadjacent energy levels when the corresponding matrix elements $x_{\mu\nu}$ are nonzero.
As illustrated on the quantum side of Fig.~\ref{fig:quantum-classical-schematic}, population can move directly from a higher to a lower energy level without passing through the intermediate levels, just as an elevator can reach a lower floor without stopping at intermediate floors.
These transitions break the locality of energy exchange underlying the classical constraints, making the Mpemba effect possible.

For a given potential and bath temperature, the bath spectrum controls the relative strengths of the allowed transitions between energy levels.
Enhancing direct transitions from high to low energy levels can reshape $\ell_2$, allowing some higher energy states to contribute less to the slowest relaxation than intermediate states.
As the initial temperature increases, the population redistributes toward higher energy levels while $\ell_2$ remains fixed.
The increased population of these states can then reduce $|a_2(\beta)|$, enabling the Mpemba effect.
Discreteness alone is insufficient to produce the Mpemba effect. 
For a harmonic oscillator, transitions occur only between adjacent levels, which leads to the absence of the Mpemba effect (see Sec.~S4 in \cite{SM} for details).

The freedom---obviously theoretical but also, to some extent, experimental---to shape the bath spectral density allows the relative transition rates to be tailored, thereby providing a route to realizing the Mpemba effect. 
We illustrate this possibility with a symmetric single-well potential, given by $V(x)=4\tanh^2(5x)+0.03x^2$.
We compare super-Ohmic bath spectra $J(\omega)=\eta\omega^s \ee^{-(\omega/\omega_c)^2}$ with $s=2$ and $s=3$ at a cutoff frequency $\omega_c$, a choice motivated by super-Ohmic phonon environments in semiconductor quantum dots~\cite{Ramsay2010Phonon,Ramsay2010Damping} and the engineering of structured baths in trapped-ion simulators~\cite{Sun2025StructuredBath}.
Here, $\eta$ sets the overall strength of the bath spectrum.
The overlap coefficient $a_{2}$ is computed by numerically diagonalizing the Pauli generator in a truncated energy eigenbasis to extract $\ell_2$, and by evaluating its overlap with the thermal initial state. 

As illustrated in the main panel of Fig.~\ref{fig:quantum-mpemba}, the two spectral exponents yield different behaviors of $a_2(\beta)$ over the range of initial temperatures we consider. 
For the parameters shown, $a_2(\beta)$ is monotonic for $s=2$, whereas it becomes nonmonotonic for $s=3$, indicating the Mpemba effect.
This comparison supports the mechanism we discussed above.
A larger spectral exponent $s$ enhances direct transitions from high to low energy levels, reducing the values of $\ell_2$ at higher energy levels.
For example, increasing $s$ from $2$ to $3$ raises $\Gamma_{0\leftarrow3}/\Gamma_{2\leftarrow3}$ from $0.81$ to $8.81$.
For $s=3$, this reduction is sufficiently strong that the thermal shift of the initial population to higher energy levels suppresses $|a_2(\beta)|$, thereby triggering the Mpemba effect.
The Mpemba effect persists over a range of large cutoff frequencies, whereas a sufficiently small $\omega_c$ can suppress it by weakening direct transitions across large energy gaps.
This result shows that quantum dynamics enables the effect even in a symmetric single-well potential, which is also forbidden in classical overdamped dynamics~\cite{liu2026mpemba,pre-us}.
The results of $a_{2}$ for the remaining three potential classes are presented in SM~\cite{SM}.

To verify the existence of the Mpemba effect for $s=3$, we numerically solve the corresponding Pauli equation~\eqref{eq:pauli} to obtain the probability distribution at different times, and use the KL divergence $D_{\mathrm{KL}}(p(t)\Vert\pi(\beta_b))=\langle\ln[p(t)/\pi(\beta_b)],p(t)\rangle$ to quantify the distance from the final distribution.
As shown in the inset of Fig.~\ref{fig:quantum-mpemba}, the hotter initial state has a larger KL divergence, but the two curves cross during relaxation, with the hotter state overtaking the colder one, confirming the Mpemba effect. 

Although the state remains diagonal in the energy basis, our model has a microscopic quantum origin and a well-defined classical limit, which distinguishes it from abstract Markov jump models. 
Quantization of the same microscopic model replaces the classical local diffusion structure with a nonlocal transition structure, which is allowed by the nonzero matrix elements $x_{\mu\nu}$.
This transition structure provides relaxation pathways that enable the Mpemba effect in regimes where the classical limit of the same microscopic model excludes it. 
Importantly, the availability of nonadjacent transitions is also a necessary condition for the thermal states to exhibit Wigner negativity~\cite{SM}, a definitive signature of quantum nature.

\begin{figure}[t]
\centering
\includegraphics[width=\columnwidth]{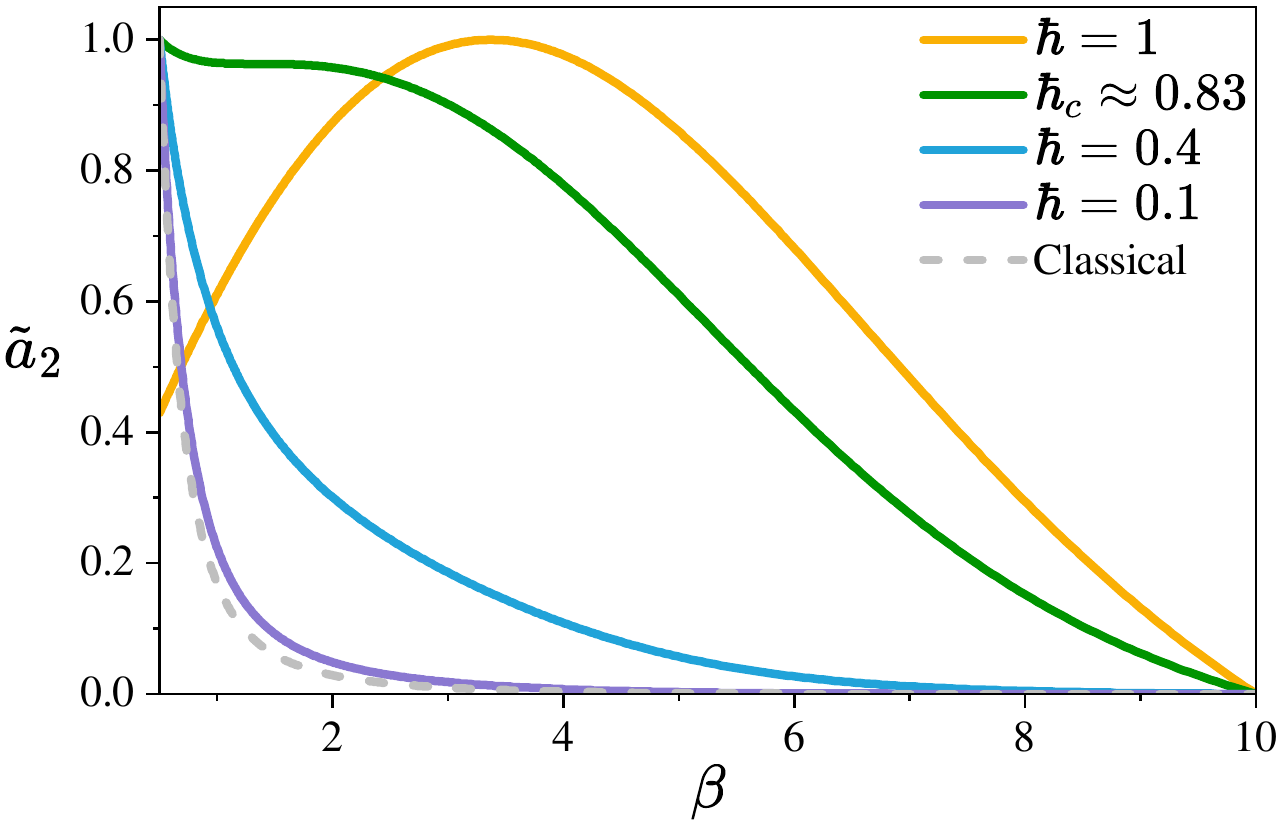}
\caption{Normalized amplitudes $\tilde a_2(\beta)$ for different values of $\hbar$.
Each curve is normalized by its own maximum over the plotted interval.
The green curve marks the threshold $\hbar_{c}\approx0.83$ for the disappearance of the Mpemba effect.
The gray dashed curve is the classical reference.
Here, $M=5$, $\eta=1$, $\beta_b=10$ and $\omega_c=8$.}
\label{fig:quantum-classical-crossover}
\end{figure}

\textit{Quantum--classical crossover.---}We now study the crossover within the Born-Markov and secular approximations underlying the Pauli equation~\eqref{eq:pauli} by decreasing the dimensionless effective Planck constant $\hbar$.
The secular approximation corresponds to averaging over the fast orbital motion on the classical side.
We use an asymmetric single-well potential $V(x)=(1-\ee^{-3x})^2+0.001x^2$ and the same super-Ohmic bath spectrum with $s=3$ as in previous examples.
The classical reference is computed independently from Eq.~\eqref{eq:classical-energy-diffusion} with the same dimensionless parameters.
As shown in Fig.~\ref{fig:quantum-classical-crossover}, $a_2(\beta)$ is nonmonotonic for $\hbar=1$ but becomes monotonic at a critical value $\hbar_{c}\approx0.83$ (depends on the specific parameters).
As $\hbar$ decreases further, the curve approaches the classical reference.
The energy levels become increasingly dense, and the Pauli dynamics approaches local energy diffusion, extinguishing the Mpemba effect.



\textit{Conclusion.---}We have shown that quantum dynamics enables the Mpemba effect in three potential classes where its classical counterpart strictly forbids it for thermal initial states.
Quantization allows for transitions between nonadjacent states, which overcomes the classical constraints and enables the Mpemba effect to occur.
As $\hbar$ decreases, local energy diffusion is recovered, restoring the classical constraints on the Mpemba effect.
The quantum--classical correspondence studied here is one of several associated with different approximations to the Caldeira--Leggett model, as outlined in the End Matter. 
Future theoretical work should explore the behavior of the Mpemba effect across these alternative dynamical regimes, but an experimental realization would certainly be the most exciting next step.

\textit{Acknowledgments.}---YL thanks Weitao Chen and Ruicheng Bao for useful discussions, and gratefully acknowledges the Yukawa Research Fellow, co-sponsored by the YITP and the Yukawa Memorial Foundation.
HH thanks Hartmut L\"{o}wen for fruitful discussions, and was supported by JSPS KAKENHI Grant No.~26K06960.
TVV was supported by JSPS KAKENHI Grant Nos.~JP23K13032, JP26K00019, JP26K00022, and JP26H02015. We thank Hartmut L\"{o}wen for sharing unpublished material with us.

\textit{Data availability.}---The data are not publicly available upon publication. The data are available from the authors upon reasonable request.

\bibliography{references}

\setcounter{equation}{0}
\renewcommand{\theequation}{E\arabic{equation}}
\renewcommand{\theHequation}{E.\arabic{equation}}
\section*{End Matter}

\begin{figure*}[t]
  \centering
  \includegraphics[width=\textwidth]{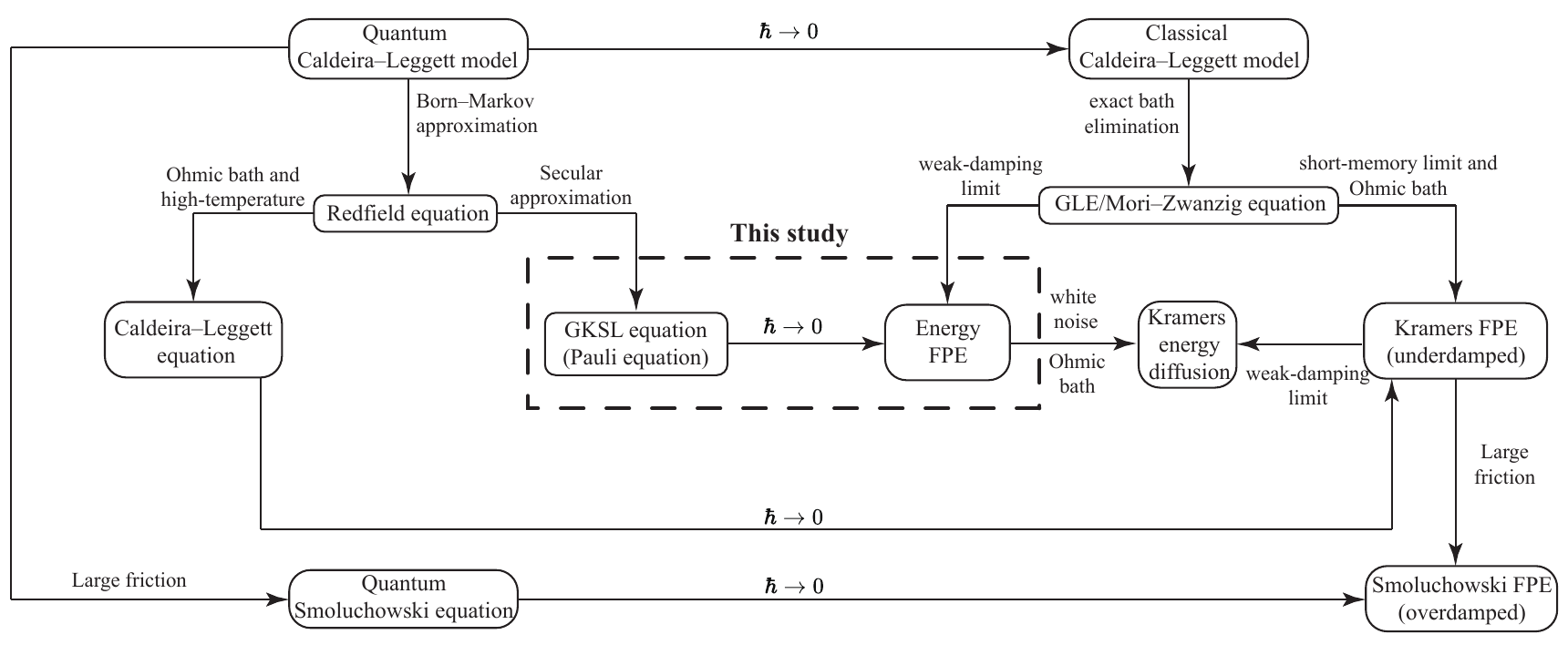}
  \caption{Reduced dynamical descriptions of the Caldeira--Leggett model.
  Solid arrows indicate approximations or limiting procedures.
  The boxed branch identifies the correspondence studied in this Letter.}
  \label{fig:dynamical-framework}
\end{figure*}

\textit{Caldeira--Leggett model.---}Figure~\ref{fig:dynamical-framework} summarizes the dynamical descriptions derived from the Caldeira--Leggett model~\cite{CaldeiraLeggett1983} using different approximations and highlights those studied in this Letter.
Notably, the energy diffusion dynamics considered here lies on a branch different  from the underdamped Kramers and overdamped Smoluchowski descriptions and cannot be regarded as an equivalent reformulation of either.
We now specify the assumptions behind each description.

The quantum Caldeira--Leggett model couples the particle to an environment of harmonic oscillators via the total Hamiltonian
\begin{equation}\label{eq:em-cl-model}
  \hat H_{\mathrm{tot}}=\hat H+\sum_j\left[\frac{\hat P_j^2}{2m_j}+\frac{m_j\omega_j^2}{2}\left(\hat Q_j-\frac{c_j\hat x}{m_j\omega_j^2}\right)^2\right],
\end{equation}
where $\hat Q_j$ and $\hat P_j$ are the position and momentum operators for the $j$-th bath oscillator, and $m_j$, $\omega_j$, and $c_j$ denote the mass, frequency, and coupling strength of bath oscillator $j$, respectively.

The corresponding classical Hamiltonian is
\begin{equation}\label{eq:em-classical-cl-model}
  H_{\mathrm{tot}}^{\mathrm{cl}}=H_c+\sum_j\left[\frac{P_j^2}{2m_j}+\frac{m_j\omega_j^2}{2}\left(Q_j-\frac{c_jx}{m_j\omega_j^2}\right)^2\right],
\end{equation}
with $H_c=p^2/(2M)+V(x)$ and canonical phase-space variables $(x,p;\{Q_j,P_j\})$. 
This Hamiltonian is exactly the Weyl symbol of the quantum Hamiltonian~\eqref{eq:em-cl-model}, $(\hat H_{\mathrm{tot}})_W=H_{\mathrm{tot}}^{\mathrm{cl}}$, where $(\cdot)_W$ denotes the Weyl transform mapping an operator $\hat A$ to its phase-space symbol $A_W$.
The correspondence between the two models extends to their dynamics.
Therefore, the quantum Caldeira--Leggett model approaches the classical Caldeira--Leggett model in the classical limit $\hbar\to0$.

\textit{GKSL and Pauli equations.---}For the quantum Caldeira--Leggett model~\eqref{eq:em-cl-model}, the system and bath are assumed to be initially factorized, with the bath prepared in a canonical equilibrium state at inverse temperature $\beta_b$.
Under weak coupling and assuming a bath correlation time much shorter than the system relaxation time, tracing out the bath within the Born--Markov approximations yields the Redfield equation.
The subsequent secular approximation averages out rapidly oscillating off-resonant terms, leading to the GKSL equation~\eqref{eq:lindblad}~\cite{Breuer2002}. 
This equation preserves the diagonal structure of a thermal initial state in the energy basis, and the populations obey the Pauli equation~\eqref{eq:pauli}.

\textit{Classical energy diffusion.---}In the classical Caldeira--Leggett model in Eq.~\eqref{eq:em-classical-cl-model}, eliminating the bath degrees of freedom gives a generalized Langevin equation~\cite{Zwanzig1973}.
After the initial bath transient, we neglect the initial-slip term associated with the factorized initial state (see SM~\cite{SM}) and write
\begin{equation}\label{eq:em-gle}
  M\ddot x+V'(x)+\int_0^t\mathcal K(t-s)\dot x(s)\,\dd s=\xi(t),
\end{equation}
where $\mathcal K$ is the friction memory kernel and $\xi$ is a zero-mean random force satisfying $\langle\xi(t)\xi(s)\rangle=\beta_b^{-1}\mathcal K(|t-s|)$.
For weak damping and a bath correlation time much shorter than the energy relaxation time, averaging over the fast orbital motion and coarse-graining over bath memory yield
Eqs.~\eqref{eq:classical-energy-diffusion} and \eqref{eq:energy-diffusion-coefficient} on each orbit branch.
These steps retain the dependence of $D_E(E)$ on the bath spectrum without requiring white noise.

The quantum Pauli equation~\eqref{eq:pauli} and the classical energy diffusion equation~\eqref{eq:classical-energy-diffusion} describe energy relaxation under a direct one-to-one correspondence between their respective approximations.
The classical weak-damping expansion corresponds to the Born approximation in retaining the leading effects of the bath coupling.
Coarse-graining over bath memory plays the role of the quantum Markov approximation, while averaging over the fast orbital motion parallels the secular approximation.

The classical energy diffusion equation also follows from the semiclassical limit of the Pauli equation~\cite{KarrleinGrabert1998} (see Sec.~(S3) in SM~\cite{SM} for details).
As $\hbar\to0$, the energy levels become denser in dimensionless units, and transitions across a fixed number $k$ of levels near energy $E$ satisfy
\begin{equation}\label{eq:em-semiclassical-correspondence}
  \frac{E_{\mu+k}-E_\mu}{\hbar}\longrightarrow k\Omega(E),\quad |x_{\mu,\mu+k}|^2\longrightarrow |x_k(E)|^2.
\end{equation}
The relation $p_\mu(t)\simeq\hbar\Omega(E_\mu)P(E_\mu,t)$ connects the discrete populations to the continuous energy density.
In this limit, a Kramers--Moyal expansion of the Pauli equation to second order in the vanishing energy jumps yields the diffusion coefficient in Eq.~\eqref{eq:energy-diffusion-coefficient}.
The detailed balance of the Pauli transition rates fixes the corresponding drift, recovering the identical energy diffusion equation~\eqref{eq:classical-energy-diffusion} as derived above from the classical Caldeira--Leggett model.

\textit{Caldeira--Leggett and Kramers equations.---}Alternatively, rather than adopting the secular approximation, one can apply a short-memory expansion to the Redfield equation with an Ohmic bath and retain the leading noise term in the high-temperature limit. 
This procedure leads to the Caldeira--Leggett master equation~\cite{Breuer2002}
\begin{equation}\label{eq:em-cl-master}
  \dot\rho=-\frac{\mathrm i}{\hbar}[\hat H,\rho]-\frac{\mathrm i\gamma_{\mathrm f}}{2\hbar}[\hat x,\{\hat p,\rho\}]-\frac{M\gamma_{\mathrm f}\beta_b^{-1}}{\hbar^2}[\hat x,[\hat x,\rho]],
\end{equation}
where $\gamma_{\mathrm f}$ is the momentum damping rate.
This equation preserves the trace and Hermiticity of $\rho$, but is not of the GKSL form and thus does not in general generate a completely positive dynamical map and may fail to preserve positivity.

On the classical side, when the memory time of an Ohmic bath is negligible compared with the timescales of orbital motion and momentum relaxation, the memory kernel in Eq.~\eqref{eq:em-gle} contracts to the Dirac $\delta$-function. 
The generalized Langevin dynamics then reduces to the standard underdamped Langevin equation with local friction and Gaussian white noise, whose phase-space distribution $F(x,p,t)$ obeys the Kramers equation~\cite{risken1996fokker}
\begin{equation}\label{eq:em-kramers}
  \begin{aligned}
    \partial_tF={}&-\frac pM\partial_xF+\partial_p[(V'+\gamma_{\mathrm f}p)F]\\
    &+M\gamma_{\mathrm f}\beta_b^{-1}\partial_p^2F.
  \end{aligned}
\end{equation}
In the Wigner representation, the quantum Caldeira--Leggett equation~\eqref{eq:em-cl-master} reduces to the Kramers equation~\eqref{eq:em-kramers} as $\hbar\to0$.
Furthermore, in the weak-damping regime, averaging Eq.~\eqref{eq:em-kramers} over the fast orbital motion gives the Kramers energy diffusion equation discussed in Ref.~\cite{Shapira2026Underdamped}.
The same equation is recovered from Eq.~\eqref{eq:classical-energy-diffusion} for an Ohmic bath in the limit of white noise.

\textit{Quantum and classical Smoluchowski equations.---}For the quantum Caldeira--Leggett model with an Ohmic bath, the strong-friction limit yields the quantum Smoluchowski equation containing the leading quantum corrections~\cite{Ankerhold2001,Maier2010},
\begin{equation}\label{eq:em-qse}
  \partial_tw_q=\frac1{M\gamma_{\mathrm f}}\partial_x\left\{V'w_q+\partial_x[D_q(x)w_q]\right\}.
\end{equation}
Here, $w_q(x,t)=\langle x|\rho(t)|x\rangle$ is the position probability density and $D_q(x)=\beta_b^{-1}[1+\beta_b\Lambda V''(x)]$, where $\Lambda$ measures the quantum correction to position fluctuations.
This approximation requires $|\beta_b\Lambda V''|\ll1$.

In the classical strong-friction regime, adiabatic elimination of momentum from the Kramers equation~\eqref{eq:em-kramers} gives the classical Smoluchowski equation~\cite{risken1996fokker}
\begin{equation}\label{eq:em-classical-smoluchowski}
  \partial_tw_c=\frac1{M\gamma_{\mathrm f}}\partial_x\left[V'w_c+\beta_b^{-1}\partial_xw_c\right],
\end{equation}
which constitutes the overdamped Fokker--Planck equation for the marginal position distribution $w_c(x,t)=\int F(x,p,t)\,\dd p$.
Taking the limit $\hbar\to0$ drives $\Lambda\to0$ and hence $D_q(x)\to\beta_b^{-1}$, under which the quantum Smoluchowski equation~\eqref{eq:em-qse} reduces to the classical Smoluchowski equation~\eqref{eq:em-classical-smoluchowski}.

\end{document}


\title{Supplemental Material for\\
``Quantization Breaks Classical Mpemba Constraints''}

\author{Yue Liu}
\affiliation{Center for Gravitational Physics and Quantum Information, Yukawa Institute for Theoretical Physics, Kyoto University, Kitashirakawa Oiwakecho, Sakyo-ku, Kyoto 606-8502, Japan}
\author{Tan Van Vu}
\affiliation{Center for Gravitational Physics and Quantum Information, Yukawa Institute for Theoretical Physics, Kyoto University, Kitashirakawa Oiwakecho, Sakyo-ku, Kyoto 606-8502, Japan}
\author{Fr\'ed\'eric van Wijland}
\affiliation{Laboratoire Mati\`ere et Syst\`emes Complexes (MSC), Universit\'e Paris Cit\'e \& CNRS (UMR 7057), 75013 Paris, France}
\affiliation{Center for Gravitational Physics and Quantum Information, Yukawa Institute for Theoretical Physics, Kyoto University, Kitashirakawa Oiwakecho, Sakyo-ku, Kyoto 606-8502, Japan}
\author{Hisao Hayakawa}
\affiliation{Center for Gravitational Physics and Quantum Information, Yukawa Institute for Theoretical Physics, Kyoto University, Kitashirakawa Oiwakecho, Sakyo-ku, Kyoto 606-8502, Japan}
\maketitle

\renewcommand{\thesection}{S\arabic{section}}
\renewcommand{\theequation}{S\arabic{equation}}
\renewcommand{\thefigure}{S\arabic{figure}}
\renewcommand{\thetable}{S\arabic{table}}
\setcounter{secnumdepth}{2}

\setcounter{tocdepth}{2}
\begingroup
\makeatletter
\let\l@subsubsection\@gobbletwo
\makeatother
\tableofcontents
\endgroup

\section{From the Caldeira--Leggett model to the Pauli equation}
\label{sec:sm-quantum-dynamics}
\subsection{Quantum Caldeira--Leggett model}
The system Hamiltonian is $\hat H=\hat p^2/(2M)+V(\hat x)$.
The particle couples through its position to a bath of harmonic oscillators. 
The total Hamiltonian is
\begin{equation}
  \hat H_{\mathrm{tot}}=\hat H+\sum_j\left[\frac{\hat P_j^2}{2m_j}+\frac{m_j\omega_j^2}{2}
  \left(\hat Q_j-\frac{c_j\hat x}{m_j\omega_j^2}\right)^2\right].
  \label{eq:sm-cl-hamiltonian}
\end{equation}
Expanding the square gives
\begin{equation}
  \begin{aligned}
    \hat H_{\mathrm{tot}}&=\hat H+\hat H_B+\hat H_I+\hat H_{\mathrm{ct}},
    &\hat H_B&=\sum_j\left(\frac{\hat P_j^2}{2m_j}
      +\frac{m_j\omega_j^2\hat Q_j^2}{2}\right),\\
    \hat H_I&=\hat x\otimes\hat B,
    &\hat B&=-\sum_j c_j\hat Q_j,
    \qquad
    \hat H_{\mathrm{ct}}=\hat x^2\sum_j\frac{c_j^2}{2m_j\omega_j^2}.
  \end{aligned}
  \label{eq:sm-cl-parts}
\end{equation}
The counterterm $\hat H_{\mathrm{ct}}$ cancels the static change of the potential caused by the bath displacement.

We assume that the system and bath are initially uncorrelated, with the bath in thermal equilibrium at inverse temperature $\beta_b$:
\begin{equation}
  \rho_{\mathrm{tot}}(0)=\rho(0)\otimes\rho_B,
  \qquad
  \rho_B=\frac{\ee^{-\beta_b\hat H_B}}{Z_B}.
  \label{eq:sm-initial-state}
\end{equation}
\subsection{Born--Markov approximations}

We use the interaction picture with respect to $\hat H_0=\hat H+\hat H_B$, with $\widetilde O(t)=\ee^{\mathrm i\hat H_0t/\hbar}O\ee^{-\mathrm i\hat H_0t/\hbar}$.
The total state obeys the von Neumann equation
\begin{equation}
  \frac{\dd \widetilde\rho_{\mathrm{tot}}(t)}{\dd t}
    =-\frac{\mathrm i}{\hbar}
      [\widetilde H_I(t)+\widetilde H_{\mathrm{ct}}(t),
       \widetilde\rho_{\mathrm{tot}}(t)].
  \label{eq:sm-von-neumann}
\end{equation}
We integrate Eq.~\eqref{eq:sm-von-neumann} once, substitute it back, and trace over the bath. The initial term linear in $\hat H_I$ vanishes because $\Tr_B(\hat B\rho_B)=0$.
Keeping terms through second order in $c_j$, including the counterterm, gives
\begin{equation}
  \begin{aligned}
    \dot{\widetilde\rho}(t)={}&
    -\frac{\mathrm i}{\hbar}
      [\widetilde H_{\mathrm{ct}}(t),\widetilde\rho(t)]\\
    &-\frac{1}{\hbar^2}\int_0^t\!\dd \tau\,
      \Tr_B\left[
      \widetilde H_I(t),
      [\widetilde H_I(t-\tau),
      \widetilde\rho_{\mathrm{tot}}(t-\tau)]\right],
  \end{aligned}
  \label{eq:sm-traced-equation}
\end{equation}
where $\widetilde\rho(t)=\Tr_B \widetilde\rho_{\mathrm{tot}}(t)$.
For weak coupling, the Born approximation keeps the bath in its equilibrium state $\rho_B$ and neglects system--bath correlations in this integral. 
We replace $\widetilde\rho_{\mathrm{tot}}(t-\tau)$ by $\widetilde\rho(t-\tau)\otimes\rho_B$ and obtain
\begin{equation}
  \begin{aligned}
    \dot{\widetilde\rho}(t)={}&
    -\frac{\mathrm i}{\hbar}
      [\widetilde H_{\mathrm{ct}}(t),\widetilde\rho(t)]\\
    &-\frac{1}{\hbar^2}\int_0^t\!\dd \tau\,
      \Tr_B\left[
      \widetilde H_I(t),
      [\widetilde H_I(t-\tau),
      \widetilde\rho(t-\tau)\otimes\rho_B]\right].
  \end{aligned}
  \label{eq:sm-born-equation}
\end{equation}
Let $\tau_B$ be the bath correlation time and $\tau_R$ the system relaxation time. When $\tau_B\ll\tau_R$, the reduced state varies negligibly over the bath correlation time. We therefore replace $\widetilde\rho(t-\tau)$ by $\widetilde\rho(t)$ in the integral:
\begin{equation}
  \begin{aligned}
    \dot{\widetilde\rho}(t)={}&
    -\frac{\mathrm i}{\hbar}
      [\widetilde H_{\mathrm{ct}}(t),\widetilde\rho(t)]\\
    &-\frac{1}{\hbar^2}\int_0^t\!\dd \tau\,
      \Tr_B\left[
      \widetilde H_I(t),
      [\widetilde H_I(t-\tau),
      \widetilde\rho(t)\otimes\rho_B]\right].
  \end{aligned}
  \label{eq:sm-time-local}
\end{equation}
For $t\gg\tau_B$, the decay of bath correlations allows us to extend the upper integration limit to infinity. 
Evaluating the bath trace then gives the Redfield equation in the Born--Markov approximation:
\begin{equation}
  \begin{aligned}
    \dot{\widetilde\rho}(t)={}&
    -\frac{\mathrm i}{\hbar}
      [\widetilde H_{\mathrm{ct}}(t),\widetilde\rho(t)]\\
    &-\frac{1}{\hbar^2}\int_0^\infty\!\dd \tau\,
    \left\{C(\tau)
      [\widetilde x(t),\widetilde x(t-\tau)\widetilde\rho(t)]
    +C(-\tau)
      [\widetilde\rho(t)\widetilde x(t-\tau),\widetilde x(t)]\right\}.
  \end{aligned}
  \label{eq:sm-born-markov}
\end{equation}
Here $C(\tau)=\Tr_B[\hat B(\tau)\hat B(0)\rho_B]$ is the bath correlation function, with $\hat B(\tau)=\ee^{\mathrm i\hat H_B\tau/\hbar}\hat B\ee^{-\mathrm i\hat H_B\tau/\hbar}$.
The bath state is stationary, leaving $C(\tau)$ depends only on the time difference.
Since $\hat B$ is Hermitian, $C(-\tau)=C(\tau)^*$.
The free evolution of $\widetilde x$ is kept inside the integral.

We now evaluate $C(\tau)$ for the bath.
Write the bath coordinate as $\hat Q_j=\sqrt{\hbar/(2m_j\omega_j)} (b_j+b_j^\dagger)$, where $b_j$ is the annihilation operator of the $j$th oscillator.
Its free evolution is $b_j(t)=b_j \ee^{-\mathrm i\omega_jt}$.
The nonzero thermal averages are
\begin{equation}
  \langle b_j^\dagger b_k\rangle_B=\delta_{jk}n_B(\omega_j),
  \qquad
  \langle b_jb_k^\dagger\rangle_B=\delta_{jk}[n_B(\omega_j)+1],
  \qquad
  n_B(\omega)=\frac{1}{\ee^{\beta_b\hbar\omega}-1}.
  \label{eq:sm-bath-averages}
\end{equation}
Here $\langle\cdot\rangle_B$ denotes the average in $\rho_B$.
The bath correlation function therefore reads
\begin{equation}
  C(\tau)=\sum_j\frac{\hbar c_j^2}{2m_j\omega_j}
    \left\{[n_B(\omega_j)+1]\ee^{-\mathrm i\omega_j\tau}
    +n_B(\omega_j)\ee^{\mathrm i\omega_j\tau}\right\}.
  \label{eq:sm-bath-correlation-sum}
\end{equation}
To write this sum as a frequency integral, we introduce the bath
spectral density, using the same convention as in the main text:
\begin{equation}
  J(\omega)=\pi\sum_j\frac{c_j^2}{m_j\omega_j}
    \delta(\omega-\omega_j),\qquad \omega>0.
  \label{eq:sm-spectral-density}
\end{equation}
The correlation function becomes
\begin{equation}
  C(\tau)=\frac{\hbar}{2\pi}\int_0^\infty\!\dd \omega'\,
    J(\omega')\left\{[n_B(\omega')+1]\ee^{-\mathrm i\omega'\tau}
    +n_B(\omega')\ee^{\mathrm i\omega'\tau}\right\}.
  \label{eq:sm-bath-correlation}
\end{equation}

\subsection{Secular approximation and GKSL equation}

To apply the secular approximation, we first make the oscillation frequencies in Eq.~\eqref{eq:sm-born-markov} explicit.
Let $\hat H|\mu\rangle=E_\mu|\mu\rangle$ and $x_{\mu\nu}=\langle \mu|\hat x|\nu\rangle$.
Expanding $\widetilde x(t)$ in the energy eigenbasis and grouping transitions with the same energy difference $E_\nu-E_\mu=\hbar\omega$ gives
\begin{equation}
  \begin{aligned}
    \widetilde x(t)
      &=\sum_{\mu,\nu}x_{\mu\nu}\ee^{-\mathrm i(E_\nu-E_\mu)t/\hbar}
        |\mu\rangle\langle \nu|
       =\sum_\omega \ee^{-\mathrm i\omega t}A_\omega,\\
    A_\omega&=\sum_{E_\nu-E_\mu=\hbar\omega}
      x_{\mu\nu}|\mu\rangle\langle \nu|.
  \end{aligned}
  \label{eq:sm-frequency-blocks}
\end{equation}
These operators satisfy $[\hat H,A_\omega]=-\hbar\omega A_\omega$ and
$A_{-\omega}=A_\omega^\dagger$. 
Positive $\omega$ therefore corresponds to energy loss by the system.
This expansion expresses the bath memory integral in terms of the one-sided Fourier transform of $C(\tau)$ at each transition frequency
$\omega=(E_\nu-E_\mu)/\hbar$:
\begin{equation}
  G(\omega)=\frac{1}{\hbar^2}\int_0^\infty\!\dd \tau\,
    \ee^{\mathrm i\omega\tau}C(\tau)
    =\frac{\gamma(\omega)}{2}+\mathrm i S(\omega).
  \label{eq:sm-one-sided-transform}
\end{equation}
Substitution into Eq.~\eqref{eq:sm-born-markov} gives
\begin{equation}
  \begin{aligned}
    \dot{\widetilde\rho}(t)={}&
    -\frac{\mathrm i}{\hbar}
      [\widetilde H_{\mathrm{ct}}(t),\widetilde\rho(t)]\\
    &+\sum_{\omega,\omega'}\ee^{\mathrm i(\omega'-\omega)t}
    \left\{G(\omega)
      [A_\omega\widetilde\rho,A_{\omega'}^\dagger]
    +G(\omega')^*
      [A_\omega,\widetilde\rho A_{\omega'}^\dagger]\right\}.
  \end{aligned}
  \label{eq:sm-nonsecular-equation}
\end{equation}
The secular approximation averages over a time $\Delta t$ such that
\begin{equation}
  \tau_B\ll\Delta t\ll\tau_R,
  \qquad
  |\omega-\omega'|\Delta t\gg1
  \quad\text{for distinct relevant frequencies }\omega\ne\omega'.
  \label{eq:sm-secular-condition}
\end{equation}
Terms with $\omega\ne\omega'$ then average to zero. 
All transitions with the same frequency remain in one $A_\omega$.
For a retained frequency, the two commutators combine to give
\begin{equation}
  \begin{aligned}
    &G(\omega)[A_\omega\rho,A_\omega^\dagger]
      +G(\omega)^*[A_\omega,\rho A_\omega^\dagger]\\
    &\qquad=\gamma(\omega)
      \left(A_\omega\rho A_\omega^\dagger
      -\frac12\{A_\omega^\dagger A_\omega,\rho\}\right)
      -\mathrm iS(\omega)[A_\omega^\dagger A_\omega,\rho].
  \end{aligned}
  \label{eq:sm-secular-block}
\end{equation}
The real coefficient is the two-sided bath transform,
\begin{equation}
  \gamma(\omega)=\frac{1}{\hbar^2}
    \int_{-\infty}^{\infty}\!\dd \tau\,
    \ee^{\mathrm i\omega\tau}C(\tau).
  \label{eq:sm-two-sided-transform}
\end{equation}
Using Eq.~\eqref{eq:sm-bath-correlation} and $\int_{-\infty}^{\infty}\ee^{\mathrm i y\tau}\dd \tau=2\pi\delta(y)$, we obtain, for $\omega>0$,
\begin{equation}
  \gamma(+\omega)=\frac{J(\omega)}{\hbar}[n_B(\omega)+1],
  \qquad
  \gamma(-\omega)=\frac{J(\omega)}{\hbar}n_B(\omega).
  \label{eq:sm-bath-rates}
\end{equation}
Both rates are nonnegative. The identity
$n_B(\omega)+1=\ee^{\beta_b\hbar\omega}n_B(\omega)$ gives
\begin{equation}
  \gamma(-\omega)=\ee^{-\beta_b\hbar\omega}\gamma(+\omega),
  \label{eq:sm-thermal-balance}
\end{equation}
which is the detailed-balance condition used in the main text.

The imaginary part produces an energy shift. 
Equivalently,
\begin{equation}
  S(\omega)=\frac{1}{2\pi}\operatorname{P}
    \int_{-\infty}^{\infty}\!\dd \omega'\,
    \frac{\gamma(\omega')}{\omega-\omega'},
  \label{eq:sm-principal-value}
\end{equation}
where $\operatorname{P}$ denotes the Cauchy principal value.
Under the secular approximation, we retain the diagonal part of the counterterm in the energy basis and include it in the Hamiltonian correction:
\begin{equation}
  \hat H_{\mathrm{LS}}
  =\sum_\nu\langle \nu|\hat H_{\mathrm{ct}}|\nu\rangle|\nu\rangle\langle \nu|
   +\hbar\sum_\omega S(\omega)A_\omega^\dagger A_\omega,
  \label{eq:sm-hamiltonian-shift}
\end{equation}
which is the Lamb shift Hamiltonian and commutes with $\hat H$.
We now return to Eq.~\eqref{eq:sm-nonsecular-equation}.
Keeping only the terms with $\omega=\omega'$ and using Eq.~\eqref{eq:sm-secular-block}, we obtain
\begin{equation*}
  \dot{\widetilde\rho}
    =-\frac{\mathrm i}{\hbar}[\hat H_{\mathrm{LS}},\widetilde\rho]
    +\sum_\omega\gamma(\omega)
    \left(A_\omega\widetilde\rho A_\omega^\dagger
    -\frac12\{A_\omega^\dagger A_\omega,\widetilde\rho\}\right),
\end{equation*}
where the Hamiltonian terms have been combined using Eq.~\eqref{eq:sm-hamiltonian-shift}.
The state in the Schr\"odinger-picture is $\rho(t)=\ee^{-\mathrm i\hat Ht/\hbar}\widetilde\rho(t)\ee^{\mathrm i\hat Ht/\hbar}$.
Differentiating this relation adds the free evolution term $-\mathrm i[\hat H,\rho]/\hbar$.
The phases of $A_\omega$ and $A_\omega^\dagger$ cancel in each dissipative term, giving the GKSL equation
\begin{equation}
  \dot\rho=-\frac{\mathrm i}{\hbar}
    [\hat H+\hat H_{\mathrm{LS}},\rho]
    +\sum_\omega\gamma(\omega)
    \left(A_\omega\rho A_\omega^\dagger
    -\frac12\{A_\omega^\dagger A_\omega,\rho\}\right).
  \label{eq:sm-lindblad}
\end{equation}
In the main text, we absorb the Lamb shift into the system Hamiltonian and write $\hat H\to\hat H+\hat H_{\mathrm{LS}}$.

\subsection{Pauli equation}

The thermal initial state at inverse temperature $\beta$ is
\begin{equation}
  \rho(0)=\sum_\mu p_\mu(0)|\mu\rangle\langle \mu|,
  \qquad
  p_\mu(0)=\frac{\ee^{-\beta E_\mu}}{Z_q(\beta)},
  \qquad
  Z_q(\beta)=\sum_\mu \ee^{-\beta E_\mu}.
  \label{eq:sm-thermal-populations}
\end{equation}
This state is diagonal in the energy basis and remains diagonal under Eq.~\eqref{eq:sm-lindblad}.
Writing $\rho(t)=\sum_\mu p_\mu(t)|\mu\rangle\langle\mu|$, we take the diagonal matrix elements of Eq.~\eqref{eq:sm-lindblad}.
The Hamiltonian commutator vanishes, leaving
\begin{equation}
  \dot p_\mu=\sum_{\omega,\nu}\gamma(\omega)
    \left[|\langle\mu|A_\omega|\nu\rangle|^2p_\nu
    -|\langle\nu|A_\omega|\mu\rangle|^2p_\mu\right].
  \label{eq:sm-population-gain-loss}
\end{equation}
By Eq.~\eqref{eq:sm-frequency-blocks}, $\langle\mu|A_\omega|\nu\rangle=x_{\mu\nu}$ when $E_\nu-E_\mu=\hbar\omega$ and is zero otherwise.
Performing the sum over $\omega$ gives the Pauli equation
\begin{equation}
  \dot p_\mu=\sum_{\nu\ne \mu}
    \left(\Gamma_{\mu\leftarrow \nu}p_\nu
    -\Gamma_{\nu\leftarrow \mu}p_\mu\right)
    \equiv(\mathcal L_q\mathbf p)_\mu,
  \qquad
  \Gamma_{\mu\leftarrow \nu}=|x_{\mu\nu}|^2
    \gamma\!\left(\frac{E_\nu-E_\mu}{\hbar}\right),
  \label{eq:sm-pauli}
\end{equation}
where $\mathbf p(t)=(p_0(t),p_1(t),\ldots)^{\mathsf T}$ is the population vector.
This is the Pauli equation in the main text. Its off-diagonal generator entries are $(\mathcal L_q)_{\mu\nu}=\Gamma_{\mu\leftarrow \nu}$, and its diagonal entries are $(\mathcal L_q)_{\mu\mu}=-\sum_{\nu\ne \mu}\Gamma_{\nu\leftarrow \mu}$.
Using $|x_{\mu\nu}|^2=|x_{\nu\mu}|^2$, Eq.~\eqref{eq:sm-thermal-balance} gives the detailed-balance relation
\begin{equation}
  \Gamma_{\nu\leftarrow\mu}
  =\ee^{-\beta_b(E_\nu-E_\mu)}\Gamma_{\mu\leftarrow\nu}.
  \label{eq:sm-population-balance}
\end{equation}
Thus, the Gibbs populations $p_\mu=\ee^{-\beta_b E_\mu}/Z_q(\beta_b)$ form a stationary solution of the Pauli equation.

\section{Classical energy diffusion from the microscopic model}
\label{sec:sm-classical-dynamics}
We give the classical Caldeira--Leggett model and its relation to the quantum model. 
We then eliminate the bath and average over the fast orbital motion to obtain the energy diffusion equation.

\subsection{Classical Caldeira--Leggett model}

The classical system and bath have canonical coordinates $(x,p;\{Q_j,P_j\})$. The Hamiltonian of the classical Caldeira--Leggett model is
\begin{equation}
  H_{\mathrm{tot}}^{\mathrm{cl}}=H_c(x,p)
  +\sum_j\left[\frac{P_j^2}{2m_j}
  +\frac{m_j\omega_j^2}{2}
  \left(Q_j-\frac{c_jx}{m_j\omega_j^2}\right)^2\right],
  \qquad H_c(x,p)=\frac{p^2}{2M}+V(x).
  \label{eq:sm-classical-hamiltonian}
\end{equation}
The equations of motion read
\begin{equation}
  M\ddot x+V'(x)
    =\sum_j c_j\left(Q_j-\frac{c_jx}{m_j\omega_j^2}\right),
  \qquad
  m_j\ddot Q_j+m_j\omega_j^2Q_j=c_jx.
  \label{eq:sm-classical-motion}
\end{equation}

Equation~\eqref{eq:sm-classical-hamiltonian} is the Weyl symbol of the quantum Hamiltonian in Eq.~\eqref{eq:sm-cl-hamiltonian}.
For smooth phase-space symbols, the commutator becomes a Poisson bracket:
\begin{equation}
  \left(\frac{[\hat A,\hat H_{\mathrm{tot}}]}{\mathrm i\hbar}\right)_W
    =\{A_W,H_{\mathrm{tot}}^{\mathrm{cl}}\}_{\mathrm{PB}}
      +\mathcal O(\hbar^2).
  \label{eq:sm-classical-bracket-limit}
\end{equation}
Here the subscript $W$ denotes a Weyl symbol, and the Poisson bracket includes both system and bath coordinates.
Thus the quantum equations for position and momentum approach Eq.~\eqref{eq:sm-classical-motion} as $\hbar\to0$.

\subsection{Generalized Langevin equation}

The bath oscillator in Eq.~\eqref{eq:sm-classical-motion} has the exact solution
\begin{equation}
  Q_j(t)=Q_j(0)\cos\omega_jt
    +\frac{P_j(0)}{m_j\omega_j}\sin\omega_jt
    +\frac{c_j}{m_j\omega_j}
      \int_0^t\!\dd s\,
      \sin[\omega_j(t-s)]x(s).
  \label{eq:sm-bath-solution}
\end{equation}
The first two terms are the free oscillation of the bath oscillator, and the last term represents the influence of the system on the bath.
Integrating the last term by parts and substituting it into the system equation gives
\begin{equation}
  M\ddot x(t)+V'[x(t)]
    +\int_0^t\!\dd s\,\mathcal K(t-s)\dot x(s)
    =\xi_0(t)-\mathcal K(t)x(0),
  \label{eq:sm-gle-with-slip}
\end{equation}
with
\begin{equation}
  \begin{aligned}
    \mathcal K(t)&=\sum_j\frac{c_j^2}{m_j\omega_j^2}\cos\omega_jt
      =\frac{1}{\pi}\int_0^\infty\!\dd \omega\,
        \frac{J(\omega)}{\omega}\cos\omega t,\\
    \xi_0(t)&=\sum_j c_j\left[
      Q_j(0)\cos\omega_jt
      +\frac{P_j(0)}{m_j\omega_j}\sin\omega_jt\right].
  \end{aligned}
  \label{eq:sm-classical-kernel-noise}
\end{equation}
The counterterm cancels the term proportional to $x(t)$, and the definition of spectral density $J(\omega)$ is the same as in the quantum model (Eq.~\eqref{eq:sm-spectral-density}).

We take the system and bath to be initially independent.
The bath is in thermal equilibrium at inverse temperature $\beta_b$ with respect to its free Hamiltonian, so its initial probability density is
\begin{equation*}
  f_B\propto\exp\!\left[-\beta_b\sum_j
    \left(\frac{P_j(0)^2}{2m_j}
    +\frac{m_j\omega_j^2Q_j(0)^2}{2}\right)\right].
\end{equation*}
The bath variables are independent and have zero mean.
Equipartition then gives
\begin{equation}
  \langle Q_j(0)Q_k(0)\rangle
    =\delta_{jk}\frac{1}{\beta_b m_j\omega_j^2},
  \qquad
  \langle P_j(0)P_k(0)\rangle
    =\delta_{jk}\frac{m_j}{\beta_b},
  \qquad
  \langle Q_j(0)P_k(0)\rangle=0.
  \label{eq:sm-classical-bath-moments}
\end{equation}
Using the noise definition in Eq.~\eqref{eq:sm-classical-kernel-noise} and the thermal averages in Eq.~\eqref{eq:sm-classical-bath-moments}, we obtain
\begin{equation}
  \langle\xi_0(t)\rangle=0,
  \qquad
  \langle\xi_0(t)\xi_0(s)\rangle
    =\beta_b^{-1}\mathcal K(t-s),
  \label{eq:sm-classical-fdt}
\end{equation}
which is the classical fluctuation--dissipation relation.

The quantity $\beta_b^{-1}\mathcal K(\tau)$ plays the role of the bath correlation function in the classical model.
To compare it with the quantum correlation $C(\tau)$, we rewrite Eq.~\eqref{eq:sm-bath-correlation} as
\begin{equation}
  C(\tau)=\frac{\hbar}{2\pi}\int_0^\infty\!\dd \omega\,
  J(\omega)\left[
    \coth\!\left(\frac{\beta_b\hbar\omega}{2}\right)\cos\omega\tau
    -\mathrm i\sin\omega\tau\right].
  \label{eq:sm-bath-coth-form}
\end{equation}
In the classical limit $\hbar\to0$, we have $\hbar\coth(\beta_b\hbar\omega/2)\to2/(\beta_b\omega)$.
The imaginary part vanishes, and
\begin{equation}
  C(\tau)\xrightarrow{\hbar\to0}
    \frac{1}{\pi\beta_b}\int_0^\infty\!\dd \omega\,
      \frac{J(\omega)}{\omega}\cos\omega\tau
    =\beta_b^{-1}\mathcal K(\tau),
  \label{eq:sm-classical-bath-limit}
\end{equation}
where $\mathcal K$ is the memory kernel in Eq.~\eqref{eq:sm-classical-kernel-noise}.
This agrees with the classical noise correlation in Eq.~\eqref{eq:sm-classical-fdt}.

We consider the dynamics after the initial transient, when $\mathcal K(t)$ has decayed and the initial-slip term $-\mathcal K(t)x(0)$ in Eq.~\eqref{eq:sm-gle-with-slip} is negligible.
Writing $\xi(t)=\xi_0(t)$ then gives the generalized Langevin equation
\begin{equation}
  M\ddot x(t)+V'[x(t)]
    +\int_0^t\!\dd s\,\mathcal K(t-s)\dot x(s)=\xi(t).
  \label{eq:sm-classical-gle}
\end{equation}

\subsection{Energy diffusion equation}

The system energy $E=M\dot x^2/2+V(x)$ obeys
\begin{equation}
  \dot E(t)=\dot x(t)\xi(t)
    -\dot x(t)\int_0^t\!\dd s\,
      \mathcal K(t-s)\dot x(s).
  \label{eq:sm-classical-energy-balance}
\end{equation}
Let $\tau_{\mathrm{orb}}(E)$ be the period of the isolated orbit at energy $E$
and $\Omega(E)=2\pi/\tau_{\mathrm{orb}}(E)$ its angular frequency.
We denote the energy relaxation time by $\tau_E$.
Under weak damping, the orbital period is much shorter than the
energy relaxation time, $\tau_{\mathrm{orb}}(E)\ll\tau_E$.
We also assume that the bath correlation time $\tau_B$ satisfies
$\tau_B\ll\tau_E$.
We can therefore choose an averaging interval $\Delta t$ satisfying
\begin{equation}
  \max\{\tau_B,\tau_{\mathrm{orb}}(E)\}
    \ll\Delta t\ll\tau_E.
  \label{eq:sm-classical-time-window}
\end{equation}
This interval is long enough to average over orbital motion and bath
fluctuations, but short enough that the energy remains approximately constant.
We expand the periodic motion at energy $E$ in a Fourier series:
\begin{equation}
  x_E(t)=\sum_{k\in\mathbb Z}x_k(E)\ee^{\mathrm i k\Omega(E)t},
  \qquad
  x_k(E)=\frac{1}{\tau_{\mathrm{orb}}(E)}
    \int_0^{\tau_{\mathrm{orb}}(E)}\!\dd t\,
    x_E(t)\ee^{-\mathrm i k\Omega(E)t}.
  \label{eq:sm-classical-orbit-fourier}
\end{equation}
Because $x_E(t)$ is real, $x_{-k}=x_k^*$.
The phase-averaged velocity correlation is
\begin{equation}
  C_v^{(E)}(u)
    =\langle\dot x_E(t+u)\dot x_E(t)\rangle_{\mathrm{orb}}
    =\sum_{k\ne0}[k\Omega(E)]^2|x_k(E)|^2
      \ee^{\mathrm i k\Omega(E)u}.
  \label{eq:sm-classical-velocity-correlation}
\end{equation}

To leading order in the coupling, the fluctuating energy increment in the interval $\Delta t$ is
\begin{equation}
  \Delta E_\xi=\int_t^{t+\Delta t}\!\dd s\,
    \dot x_E(s)\xi(s).
  \label{eq:sm-classical-energy-increment}
\end{equation}
The diffusion coefficient is defined by
$\operatorname{Var}(\Delta E_\xi)\simeq2D_E(E)\Delta t$.
To evaluate this variance, we use the two-sided noise spectrum
obtained from Eqs.~\eqref{eq:sm-classical-kernel-noise}
and~\eqref{eq:sm-classical-fdt}:
\begin{equation}
  S_\xi(\omega)=\int_{-\infty}^{\infty}\!\dd u\,
    \ee^{\mathrm i\omega u}\langle\xi(u)\xi(0)\rangle
    =\beta_b^{-1}\frac{J(|\omega|)}{|\omega|},
    \qquad\omega\ne0.
  \label{eq:sm-classical-noise-spectrum}
\end{equation}
The double time integral of the noise covariance therefore gives
\begin{equation}
  \begin{aligned}
    D_E(E)
      &=\frac12\int_{-\infty}^{\infty}\!\dd u\,
        C_v^{(E)}(u)\langle\xi(u)\xi(0)\rangle\\
      &=\frac12\sum_{k\ne0}[k\Omega(E)]^2|x_k(E)|^2
        S_\xi(k\Omega(E))\\
      &=\beta_b^{-1}\sum_{k=1}^{\infty}
        k\Omega(E)J\!\left(k\Omega(E)\right)|x_k(E)|^2.
  \end{aligned}
  \label{eq:sm-classical-diffusion-coefficient}
\end{equation}
The equal contributions from $k$ and $-k$ cancel the factor $1/2$. 
The bath spectrum is still sampled at each orbital harmonic and no white-noise approximation is needed.

The diffusion coefficient $D_E(E)$ describes energy fluctuations.
To obtain the evolution equation of the energy distribution, we also need the drift coefficient $A_E(E)$, the mean energy change per unit time.
We determine it from the equilibrium energy distribution and the requirement of zero probability current at equilibrium.

We first find the equilibrium energy distribution.
The bath integrals in the canonical distribution of Eq.~\eqref{eq:sm-classical-hamiltonian} are Gaussian.
Their values do not depend on the shifts $c_jx/(m_j\omega_j^2)$, so integrating out the bath leaves a system density proportional to $\ee^{-\beta_b H_c(x,p)}$.
The corresponding energy density is proportional to $g(E)\ee^{-\beta_b E}$, where $g(E)$ is the density of states
\begin{equation}
  g(E)=\int\!\dd x\,\dd p\,
    \delta(E-H_c(x,p))
    =\tau_{\mathrm{orb}}(E)=\frac{2\pi}{\Omega(E)}.
  \label{eq:sm-classical-density-of-states}
\end{equation}
The second equality follows by integrating over momentum and then over one complete orbit. The normalized equilibrium probability density is
\begin{equation}
  \pi(E;\beta_b)=\frac{g(E)\ee^{-\beta_b E}}{Z_c(\beta_b)},
  \qquad
  Z_c(\beta_b)=\int\!\dd E\,g(E)\ee^{-\beta_b E}.
  \label{eq:sm-classical-equilibrium}
\end{equation}

Let $P(E,t)\,\dd E$ be the probability in an energy interval.
A second-order Kramers--Moyal expansion gives the Fokker--Planck equation
\begin{equation}
  \partial_tP=-\partial_E(A_EP)+\partial_E^2(D_EP)
    =-\partial_E\mathcal J_E.
  \label{eq:sm-classical-fokker-planck}
\end{equation}
Here $D_E$ is already known from Eq.~\eqref{eq:sm-classical-diffusion-coefficient}, while $A_E$ remains to be determined. At equilibrium, $P(E,t)=\pi(E;\beta_b)$, and thermal detailed balance requires $A_E\pi(E;\beta_b)=\partial_E[D_E\pi(E;\beta_b)]$, giving
\begin{equation}
  A_E(E)=\frac{\partial_E[D_E(E)\pi(E;\beta_b)]}{\pi(E;\beta_b)}
    =D_E'(E)-D_E(E)
      \left[\beta_b+\partial_E\ln\Omega(E)\right].
  \label{eq:sm-classical-full-drift}
\end{equation}
The term $-\beta_bD_E$ describes energy dissipation due to friction.
Substituting the drift into Eq.~\eqref{eq:sm-classical-fokker-planck} yields
\begin{equation}
  \partial_tP(E,t)
    =\partial_E\left[
      D_E(E)\pi(E;\beta_b)
      \partial_E\left(\frac{P(E,t)}{\pi(E;\beta_b)}\right)\right]
    \equiv(\mathcal L_cP)(E,t).
  \label{eq:sm-classical-energy-diffusion}
\end{equation}
Together with Eq.~\eqref{eq:sm-classical-diffusion-coefficient}, this is the classical equation used in the main text.

\subsection{Correspondence of the approximations}

The quantum and classical derivations use corresponding approximations, as summarized in Table~\ref{tab:sm-quantum-classical-approximations}.
The Born approximation corresponds to weak damping, and the Markov approximation to coarse graining over bath memory. 
The secular approximation and orbital averaging both average over fast motion, leaving the slow evolution of the energy distribution.
This correspondence also extends to the resulting equations.
Sec.~\ref{sec:sm-semiclassical-limit} shows how the Pauli equation reduces to the classical energy-diffusion equation as $\hbar\to0$.

\begin{table}[htb]
  \centering
  \renewcommand{\arraystretch}{1.18}
  \begin{tabular}{@{}p{0.20\textwidth}p{0.36\textwidth}p{0.36\textwidth}@{}}
    \hline\hline
    \raggedright & \raggedright Quantum & \raggedright Classical \tabularnewline
    \hline
    \raggedright Weak coupling & \raggedright Born approximation & \raggedright Weak-damping approximation \tabularnewline
    \raggedright Short bath memory & \raggedright Markov approximation
                & \raggedright Coarse graining \tabularnewline
    \raggedright Fast motion & \raggedright Secular approximation & \raggedright Orbital averaging \tabularnewline
    \raggedright Thermal bath & \raggedright Detailed balance & \raggedright Fluctuation--dissipation relation \tabularnewline
    \raggedright Bath spectrum
      & \raggedright $J(|E_\nu-E_\mu|/\hbar)$
      & \raggedright $J(k\Omega(E))$ \tabularnewline
    \raggedright Evolution equation & \raggedright Pauli equation & \raggedright Energy diffusion equation \tabularnewline
    \hline\hline
  \end{tabular}
  \caption{Quantum and classical reductions of the Caldeira--Leggett
  model. Entries in the same row play analogous roles.}
  \label{tab:sm-quantum-classical-approximations}
\end{table}

\section{From the Pauli equation to classical energy diffusion}
\label{sec:sm-semiclassical-limit}

\subsection{Semiclassical correspondence}

We now derive the classical energy diffusion equation from the Pauli equation~\eqref{eq:sm-pauli}.
We want to relate the quantum energy gaps and position matrix elements to the frequency and Fourier coefficients of the classical motion.
For this purpose, we use action--angle variables.
For an orbit between the turning points $x_-(E)$ and $x_+(E)$, we define the action
\begin{equation}
  I(E)=\frac{1}{2\pi}\oint p\,\dd x
    =\frac{1}{\pi}\int_{x_-(E)}^{x_+(E)}\!\dd x\,
      \sqrt{2M[E-V(x)]},
  \qquad
  \frac{\dd I}{\dd E}=\frac{1}{\Omega(E)}.
  \label{eq:sm-semiclassical-action}
\end{equation}
The derivative follows from $\dot x=p/M$ and the fact that the motion between the two turning points takes half an orbital period.
We write $E=E(I)$ and $\Omega(I)=\Omega(E(I))$.
The action $I$ labels the orbit, while its conjugate angle $\theta$ describes the phase of motion along that orbit.
Since $H_c=E(I)$, the equations of motion give
\begin{equation}
  \dot I=0,
  \qquad
  \dot\theta=\frac{\dd E}{\dd I}=\Omega(I).
  \label{eq:sm-action-angle-motion}
\end{equation}
Choosing $\theta=0$ at $t=0$, we have $\theta=\Omega(I)t$ modulo $2\pi$.
Thus the angle increases by $2\pi$ over one orbital period.
The position is a periodic function $x(I,\theta)$, whose Fourier coefficients are $x_k(I)=x_k(E(I))$ as defined in Eq.~\eqref{eq:sm-classical-orbit-fourier}.

To relate the action to the quantum levels, we use the WKB approximation.
We match the oscillating wave function between the two regular turning points to the decaying solutions outside.
The phase corrections at the two turning points give the Bohr--Sommerfeld quantization condition for the action,
\begin{equation}
  I(E_\mu)=I_\mu+\mathcal O(\hbar^2),
  \qquad
  I_\mu=\hbar\left(\mu+\frac12\right).
  \label{eq:sm-action-quantization}
\end{equation}
Since $I$ and $\theta$ are canonically conjugate, the action operator in the semiclassical angle representation has the differential part $-\mathrm i\hbar\partial_\theta$.
For $2\pi$-periodic wave functions, this differential operator has eigenvalues $\hbar\mu$.
Including the offset $\hbar/2$ required by
Eq.~\eqref{eq:sm-action-quantization} gives
\begin{equation}
  \hat I\simeq\hbar\left(-\mathrm i\partial_\theta+\frac12\right).
  \label{eq:sm-action-operator}
\end{equation}
The eigenvalue equation with eigenvalue $I_\mu$ then reduces to
$-\mathrm i\partial_\theta\langle\theta|\mu\rangle
=\mu\langle\theta|\mu\rangle$.
Normalizing the wave function over one angular period gives
\begin{equation}
  \langle\theta|\mu\rangle\simeq
    \frac{\ee^{\mathrm i\mu\theta}}{\sqrt{2\pi}}.
  \label{eq:sm-angle-wavefunction}
\end{equation}

We then obtain the energy gaps from the action spacing.
For two levels $\mu$ and $\nu=\mu+k$, with fixed $k>0$, the action increment is $k\hbar$.
Using $\dd E/\dd I=\Omega(I)$, Taylor expansion gives
\begin{equation}
  E_\nu-E_\mu=k\hbar\Omega(I_\mu)+\mathcal O(\hbar^2),
  \qquad
  \frac{E_\nu-E_\mu}{\hbar}\longrightarrow k\Omega(I).
  \label{eq:sm-semiclassical-gaps}
\end{equation}
We next relate the position matrix elements to the Fourier coefficients.
We use the leading semiclassical formula of matrix elements with symmetric ordering, which evaluates the classical position $x(I,\theta)$ at the mean action $\bar I_{\mu\nu}=(I_\mu+I_\nu)/2$.
Substituting the angle wave functions gives
\begin{equation}
  \begin{aligned}
    x_{\nu\mu}
      &\simeq\frac{1}{2\pi}\int_0^{2\pi}\!\dd \theta\,
        \ee^{-\mathrm i\nu\theta}x(\bar I_{\mu\nu},\theta)
        \ee^{\mathrm i\mu\theta}\\
      &=\frac{1}{2\pi}\int_0^{2\pi}\!\dd \theta\,
        x(\bar I_{\mu\nu},\theta)\ee^{-\mathrm i k\theta}
        =x_k(\bar I_{\mu\nu}).
  \end{aligned}
  \label{eq:sm-position-fourier-matrix-element}
\end{equation}
At fixed $k$, the mean action tends to $I$ as $\hbar\to0$. Since $x(I,\theta)$ is real, $x_{-k}(I)=x_k(I)^*$, so both transition directions give
\begin{equation}
  |x_{\mu\pm k,\mu}|^2\longrightarrow |x_k(I)|^2.
  \label{eq:sm-matrix-element-correspondence}
\end{equation}

\subsection{Diffusion limit in action space}

Let $\Gamma_k(I_\mu)=\Gamma_{\mu+k\leftarrow\mu}$ be the rate of a jump from $\mu$ to $\mu+k$.
Reindexing the gain term in Eq.~\eqref{eq:sm-pauli} gives
\begin{equation}
  \dot p_\mu=\sum_{k\ne0}
    [\Gamma_k(I_\mu-k\hbar)p_{\mu-k}-\Gamma_k(I_\mu)p_\mu].
  \label{eq:sm-action-lattice-pauli}
\end{equation}
Since adjacent action coordinates differ by $\hbar$, the probability density per unit action is $P_I(I_\mu,t)=p_\mu(t)/\hbar$.
We use a smooth interpolation in $I$ and divide Eq.~\eqref{eq:sm-action-lattice-pauli} by $\hbar$.
A second-order Kramers--Moyal expansion in $k\hbar$ then gives
\begin{equation}
  \begin{aligned}
    \partial_tP_I
      &\simeq-\partial_I(A_I^{(\hbar)}P_I)
        +\partial_I^2(D_I^{(\hbar)}P_I),\\
    A_I^{(\hbar)}(I)&=\sum_{k\ne0}k\hbar\,\Gamma_k(I),
    \qquad
    D_I^{(\hbar)}(I)=\frac12\sum_{k\ne0}(k\hbar)^2\Gamma_k(I).
  \end{aligned}
  \label{eq:sm-action-drift-diffusion}
\end{equation}
We first calculate the diffusion coefficient.
Combining Eq.~\eqref{eq:sm-bath-rates} with the semiclassical relations above, and using $2n_B(\omega)+1=\coth(\beta_b\hbar\omega/2)$, gives
\begin{equation}
  \begin{aligned}
    D_I^{(\hbar)}(I)
      &=\frac12\sum_{k=1}^{\infty}(k\hbar)^2
        [\Gamma_k(I)+\Gamma_{-k}(I)]\\
      &=\frac{\hbar}{2}\sum_{k=1}^{\infty}
        k^2|x_k(I)|^2J(k\Omega(I))
        \coth\!\left(\frac{\beta_b\hbar k\Omega(I)}{2}\right)+o(1).
  \end{aligned}
  \label{eq:sm-action-diffusion-before-limit}
\end{equation}
Since $\hbar\coth(\beta_b\hbar k\Omega/2)\to2/(\beta_b k\Omega)$, we obtain
\begin{equation}
  D_I(I)=\lim_{\hbar\to0}D_I^{(\hbar)}(I)
    =\frac{1}{\beta_b\Omega(I)}
      \sum_{k=1}^{\infty}kJ(k\Omega(I))|x_k(I)|^2.
  \label{eq:sm-action-diffusion-limit}
\end{equation}

We now determine the drift from detailed balance.
The Gibbs populations give the equilibrium density in action space,
\begin{equation}
  \pi_I^{(\hbar)}(I_\mu)=\frac{\ee^{-\beta_b E_\mu}}{\hbar Z_q(\beta_b)}
    \longrightarrow
  \pi_I(I)=\frac{2\pi\ee^{-\beta_b E(I)}}{Z_c(\beta_b)}.
  \label{eq:sm-action-equilibrium-limit}
\end{equation}
Here $\hbar Z_q(\beta_b)\to\int\!\dd I\,\ee^{-\beta_b E(I)}=Z_c(\beta_b)/(2\pi)$, since $\dd I=\dd E/\Omega(E)=g(E)\dd E/(2\pi)$.
For jumps between $I$ and $I+k\hbar$, the detailed balance relation in Eq.~\eqref{eq:sm-population-balance} becomes
\begin{equation}
  \Gamma_k(I)\pi_I^{(\hbar)}(I)
    =\Gamma_{-k}(I+k\hbar)\pi_I^{(\hbar)}(I+k\hbar).
  \label{eq:sm-action-edge-balance}
\end{equation}
Using this relation in the first jump moment and expanding in $k\hbar$ gives
\begin{equation}
  \begin{aligned}
    A_I^{(\hbar)}\pi_I^{(\hbar)}
      &=\sum_{k=1}^{\infty}k\hbar
        [\Gamma_k(I)\pi_I^{(\hbar)}(I)
        -\Gamma_k(I-k\hbar)\pi_I^{(\hbar)}(I-k\hbar)]\\
      &=\partial_I\left[\sum_{k=1}^{\infty}(k\hbar)^2
        \Gamma_k(I)\pi_I^{(\hbar)}(I)\right]+o(1)\\
      &=\partial_I(D_I^{(\hbar)}\pi_I^{(\hbar)})+o(1).
  \end{aligned}
  \label{eq:sm-edge-moment-identities}
\end{equation}
Taking $\hbar\to0$ therefore gives
\begin{equation}
  A_I\pi_I=\partial_I(D_I\pi_I),
  \qquad
  A_I(I)=\frac{\dd D_I}{\dd I}
    -\beta_b\Omega(I)D_I(I).
  \label{eq:sm-action-drift-limit}
\end{equation}
Substituting this drift into Eq.~\eqref{eq:sm-action-drift-diffusion} and taking the limit gives
\begin{equation}
  \partial_tP_I
    =\partial_I\left[D_I(I)\pi_I(I)
      \partial_I\left(\frac{P_I}{\pi_I}\right)\right].
  \label{eq:sm-action-diffusion-equation}
\end{equation}

\subsection{Energy diffusion equation}

To express this result in energy, we use probability conservation, $P(E,t)\,\dd E=P_I(I,t)\,\dd I$. Since $\dd E=\Omega(I)\,\dd I$, the densities satisfy
\begin{equation}
  P(E,t)=\frac{P_I(I(E),t)}{\Omega(E)},
  \qquad
  \pi(E;\beta_b)=\frac{\pi_I(I(E))}{\Omega(E)}
    =\frac{g(E)\ee^{-\beta_b E}}{Z_c(\beta_b)}.
  \label{eq:sm-action-energy-density}
\end{equation}
Using these relations and $\partial_I=\Omega\partial_E$ in Eq.~\eqref{eq:sm-action-diffusion-equation} gives
\begin{equation}
  \partial_tP(E,t)
    =\partial_E\left[D_E(E)\pi(E;\beta_b)
      \partial_E\left(\frac{P(E,t)}{\pi(E;\beta_b)}\right)\right],
  \label{eq:sm-pauli-energy-diffusion}
\end{equation}
where
\begin{equation}
  D_E(E)=\Omega(E)^2D_I(I(E))
    =\beta_b^{-1}\sum_{k=1}^{\infty}
      k\Omega(E)J(k\Omega(E))|x_k(E)|^2.
  \label{eq:sm-pauli-energy-diffusion-coefficient}
\end{equation}
These expressions agree with Eqs.~\eqref{eq:sm-classical-energy-diffusion} and~\eqref{eq:sm-classical-diffusion-coefficient}, obtained from the classical Caldeira--Leggett model in Sec.~\ref{sec:sm-classical-dynamics}.

\section{Harmonic oscillator: an analytic reference}
\label{sec:sm-harmonic-oscillator}
Here, we consider the harmonic oscillator as an exactly solvable example of quantum case, which forbids the Mpemba effect. 
The Hamiltonian of the quantum harmonic oscillator and position operator are
\begin{equation}
  \hat H=\frac{\hat p^2}{2M}+\frac{M\Omega^2\hat x^2}{2}
    =\hbar\Omega\left(\hat a^\dagger\hat a+\frac12\right),
  \qquad
  \hat x=\sqrt{\frac{\hbar}{2M\Omega}}(\hat a+\hat a^\dagger).
  \label{eq:sm-ho-hamiltonian}
\end{equation}
The energies are $E_\mu=\hbar\Omega(\mu+1/2)$, with $\mu=0,1,\ldots$. 
Position coupling only allows transitions between adjacent levels. 
All downward transitions have the same frequency $\Omega$ ($-\Omega$ for upward transitions), so Eq.~\eqref{eq:sm-frequency-blocks} gives
\begin{equation}
  A_{+\Omega}=\sqrt{\frac{\hbar}{2M\Omega}}\hat a,
  \qquad A_{-\Omega}=\sqrt{\frac{\hbar}{2M\Omega}}\hat a^\dagger.
  \label{eq:sm-ho-jump-operators}
\end{equation}
Define the mean thermal occupation and damping rate by
\begin{equation}
  \bar n(\beta)=\frac{1}{\ee^{\beta\hbar\Omega}-1},
  \qquad
  \bar n_b=\bar n(\beta_b)=n_B(\Omega),
  \qquad
  \kappa=\frac{J(\Omega)}{2M\Omega}.
  \label{eq:sm-ho-bath-parameters}
\end{equation}
The rate coefficients in Eq.~\eqref{eq:sm-bath-rates} become
\begin{equation}
  \Gamma_\downarrow=\frac{\hbar}{2M\Omega}\gamma(+\Omega)=\kappa(\bar n_b+1),
  \qquad
  \Gamma_\uparrow=\frac{\hbar}{2M\Omega}\gamma(-\Omega)=\kappa\bar n_b.
  \label{eq:sm-ho-rate-coefficients}
\end{equation}
Thus the transition rates are $\Gamma_{\mu-1\leftarrow\mu}=\mu\Gamma_\downarrow$ and $\Gamma_{\mu+1\leftarrow\mu}=(\mu+1)\Gamma_\uparrow$.
The Pauli equation for the populations $p_\mu$ is then
\begin{equation}
  \dot p_\mu=\Gamma_\downarrow[(\mu+1)p_{\mu+1}-\mu p_\mu]
    +\Gamma_\uparrow[\mu p_{\mu-1}-(\mu+1)p_\mu],
  \qquad p_{-1}=0.
  \label{eq:sm-ho-pauli}
\end{equation}
At a given inverse temperature $\beta$, the Gibbs populations are
\begin{equation}
  \pi_\mu(\beta)=\frac{\ee^{-\beta E_\mu}}{Z_q(\beta)}
    =(1-\ee^{-\beta\hbar\Omega})\ee^{-\beta\hbar\Omega\mu}.
  \label{eq:sm-ho-equilibrium}
\end{equation}

To calculate $a_2$, we first determine the slowest left mode.
Its eigenvalue equation is
\begin{equation}
  \begin{aligned}
    (\mathcal L_q^\dagger\ell_2)_\mu
      &=\Gamma_\downarrow\mu(\ell_{2,\mu-1}-\ell_{2,\mu})\\
      &\quad+\Gamma_\uparrow(\mu+1)(\ell_{2,\mu+1}-\ell_{2,\mu})
      =-\lambda_2\ell_{2,\mu}.
  \end{aligned}
  \label{eq:sm-ho-adjoint}
\end{equation}
Taking the unnormalized form $\ell_{2,\mu}=\mu-\bar n_b$ gives
\begin{equation}
  (\mathcal L_q^\dagger\ell_2)_\mu
    =-\Gamma_\downarrow\mu+\Gamma_\uparrow(\mu+1)
    =-\kappa(\mu-\bar n_b)=-\kappa\ell_{2,\mu},
  \label{eq:sm-ho-centered-mode}
\end{equation}
which gives the slowest eigenvalue $\lambda_2=\kappa$. Detailed balance gives the right mode as
$r_{2,\mu}=\pi_\mu(\beta_b)\ell_{2,\mu}$.
The normalization condition
$\sum_{\mu=0}^\infty\ell_{2,\mu}r_{2,\mu}
=\sum_{\mu=0}^\infty\pi_\mu(\beta_b)\ell_{2,\mu}^2=1$
then gives
\begin{equation}
    \ell_{2,\mu}=\frac{\mu-\bar n_b}
        {\left[\sum_{\nu=0}^\infty(\nu-\bar n_b)^2\pi_\nu(\beta_b)\right]^{1/2}}
        =\frac{\mu-\bar n_b}{\sqrt{\bar n_b(1+\bar n_b)}}.
  \label{eq:sm-ho-normalized-modes}
\end{equation}

For a thermal initial state at inverse temperature $\beta$, the slowest overlap coefficient is
\begin{equation}
  a_2(\beta)=\sum_{\mu=0}^\infty\ell_{2,\mu}\pi_\mu(\beta)
    =\frac{\bar n(\beta)-\bar n_b}{\sqrt{\bar n_b(1+\bar n_b)}}.
  \label{eq:sm-ho-thermal-overlap}
\end{equation}
Its derivative is
\begin{equation}
  \frac{\dd a_2}{\dd \beta}
    =-\hbar\Omega
      \frac{\bar n(\beta)[1+\bar n(\beta)]}
        {\sqrt{\bar n_b(1+\bar n_b)}}<0,
  \label{eq:sm-ho-overlap-derivative}
\end{equation}
which suggests that $|a_2(\beta)|$ is monotonic in $\beta$ and the absence of the Mpemba effect.

\section{Wigner negativity and nonadjacent transitions}
\label{sec:sm-wigner-transitions}

In this section, we show that if the Wigner function takes negative values in the thermal state, the position operator must connect nonadjacent energy levels. 
Specifically,
\begin{equation}
W_\beta(x,p)<0\ \text{for some }(x,p)\quad\Longrightarrow\quad\exists\,\mu,\nu:\ |\mu-\nu|\ge2,\quad x_{\mu\nu}\ne0,
\end{equation}
where $W_\beta(x,p)$ is the Wigner function of the thermal state at inverse temperature $\beta$.

To prove this, we consider its contrapositive: if the position operator only connects adjacent energy levels, then the Wigner function is nonnegative.
Supposing $x_{\mu\nu}=0$ for all $|\mu-\nu|\ge2$, completeness of the energy eigenstates gives
\begin{equation}
  (x-x_{00})\psi_0(x)=x_{10}\psi_1(x),
\end{equation}
where $\psi_\mu(x)=\langle x|\mu\rangle$ is the energy eigenfunction in position space.
Applying $\hat H-E_0$ to this relation yields
\begin{equation}
  -\frac{\hbar^2}{M}\frac{\dd \psi_0}{\dd x}=(E_1-E_0)(x-x_{00})\psi_0,
\end{equation}
whose solution is $\psi_0(x)\propto\exp[-(E_1-E_0)M(x-x_{00})^2/(2\hbar^2)]$.
Substituting this solution into the stationary Schr\"odinger equation gives
\begin{equation}
  V(x)=E_0-\frac{E_1-E_0}{2}+\frac{(E_1-E_0)^2M}{2\hbar^2}(x-x_{00})^2,
\end{equation}
which is a harmonic potential. 
The thermal state of a harmonic oscillator is Gaussian, and its Wigner function is nonnegative. 
This proves that Wigner negativity requires nonadjacent transitions to be allowed by the position matrix elements.

\section{Additional quantum and classical examples}
\label{sec:sm-additional-examples}

The formal semiclassical correspondence is defined by $\hbar\to0$ at fixed dimensionful  classical parameters. 
Experimentally, the same decrease of the dimensionless quantum parameter $\tilde{\hbar}$ can instead be emulated by increasing the characteristic action of the physical system while keeping $\hbar$ fixed.

For numerical calculations, we choose a fixed mass unit $M_0$ (e.g., the proton mass) and a fixed time unit $t_0$.
We set $t_0$ proportional to the inverse physical energy gap between the two lowest energy levels.
The length unit $L_0$ is set by the spatial scale of the potential (e.g., the distance between two wells or the well width at a chosen reference energy).
These units define the energy scale $E_0=M_0L_0^2/t_0^2$.
We define the dimensionless variables
\begin{equation}
  \begin{aligned}
    \tilde x&=\frac{x}{L_0}, &
    \tilde E&=\frac{E}{E_0}, &
    \tilde M&=\frac{M}{M_0}, &
    \tilde t&=\frac{t}{t_0},\\
    \tilde\beta&=\beta E_0, &
    \tilde\omega&=\omega t_0, &
    \tilde{\hbar}&=\frac{\hbar}{E_0t_0}.
  \end{aligned}
  \label{eq:sm-dimensionless-variables}
\end{equation}
Here $\hbar$ is the physical reduced Planck constant, which remains fixed.
The dimensionless parameter $\tilde{\hbar}$ measures its magnitude relative to the characteristic action $E_0t_0$.
For example, substituting $x=L_0\tilde x$ into the system Hamiltonian gives
\begin{equation}
  \frac{\hat H}{E_0}
  =-\frac{\tilde{\hbar}^{2}}{2\tilde M}
    \frac{\dd ^2}{\dd \tilde x^2}
    +\tilde V(\tilde x),
  \qquad
  \tilde V(\tilde x)=\frac{V(L_0\tilde x)}{E_0},
  \label{eq:sm-dimensionless-hamiltonian}
\end{equation}
With the convention in Eq.~\eqref{eq:sm-spectral-density}, $J(\omega)$ has units of energy per length squared.
The dimensionless bath spectrum is therefore
\begin{equation}
  \tilde J(\tilde\omega)
  =\frac{L_0^2}{E_0}J\!\left(\frac{\tilde\omega}{t_0}\right)
  =\frac{t_0^2}{M_0}J\!\left(\frac{\tilde\omega}{t_0}\right).
  \label{eq:sm-dimensionless-spectral-density}
\end{equation}
For $J(\omega)=\eta\omega^s\exp[-(\omega/\omega_c)^2]$, the dimensionless spectral parameters are $\tilde\eta=\eta t_0^{2-s}/M_0$ and $\tilde\omega_c=\omega_ct_0$.
We use the same $\tilde\eta$ and $\tilde\omega_c$ when comparing the $s=2$ and $s=3$ spectra, corresponding to equal spectral densities at the reference frequency $\tilde\omega=1$.

We vary $\tilde{\hbar}$ by changing the length and energy scales of the physical potential while keeping its dimensionless form $\tilde V$ fixed.
With $M_0$ and $t_0$ fixed, increasing $L_0$ increases $E_0$ and reduces $\tilde{\hbar}$.
The physical inverse temperatures of the bath and all initial states are adjusted to keep their dimensionless values fixed.
The physical mass and bath spectrum remain unchanged, preserving the corresponding dimensionless parameters.
For simplicity, we omit the tildes in the main text and in the numerical examples below.

\subsection{Quantum examples for four potential classes}
\label{sec:sm-quantum-examples}

\begin{figure}[tbp]
\centering
\includegraphics[width=0.8\linewidth]{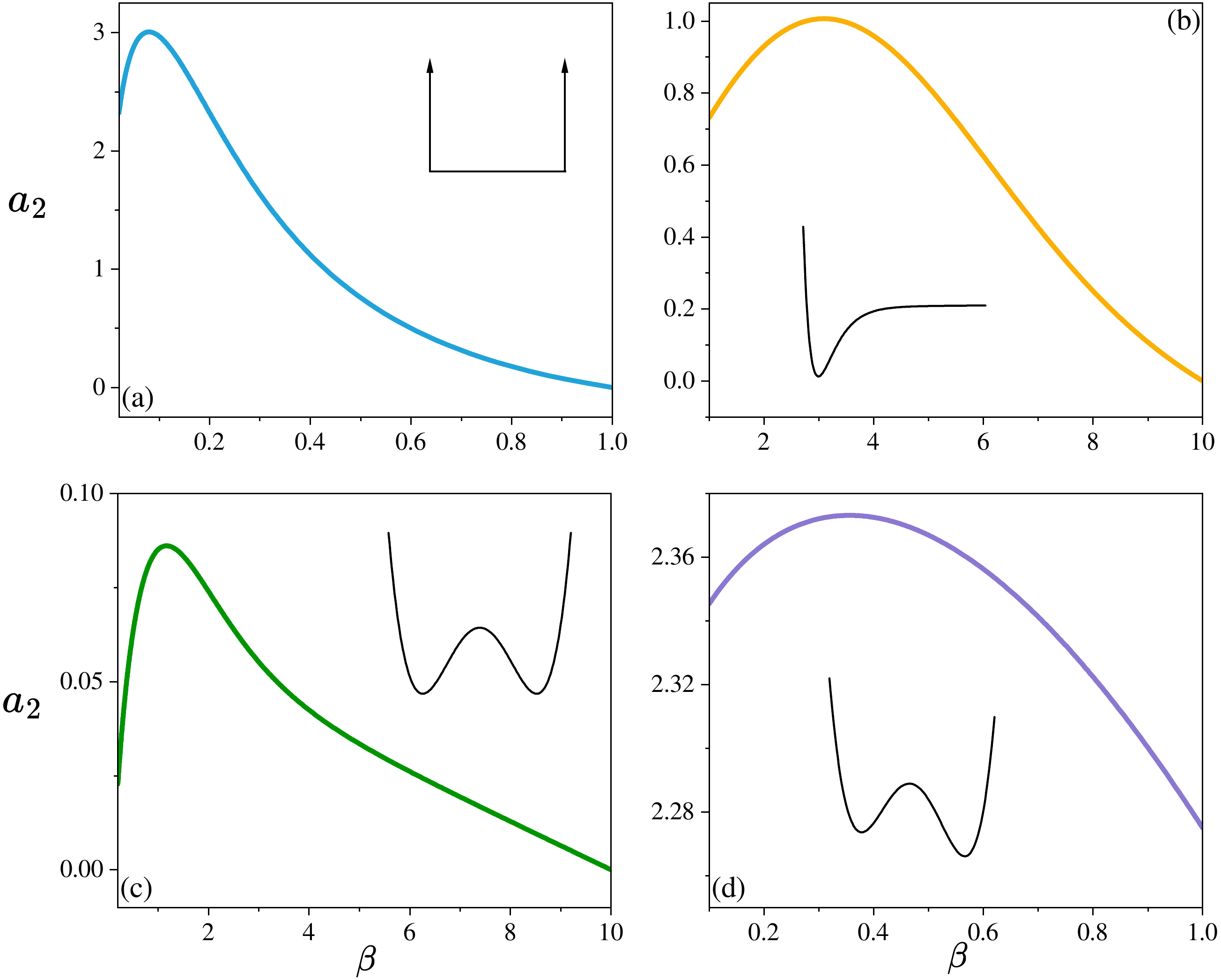}
\caption{The overlap coefficient $a_2(\beta)$ for (a) infinite square well, (b) asymmetric single well, (c) symmetric double well, and (d) asymmetric double well. The insets sketch the potentials without coordinate scales.
The masses are $M=\pi^2/2$ in (a), $M=6$ in (b) and (c), and $M=17$ in (d).
The bath inverse temperature is $\beta_b=1$ in (a), and $\beta_b=10$ and $\omega_c=8$ in (b)--(d). 
We set $\eta=1$ in all panels and $L=3$ for the double-well potentials in (c) and (d).}
\label{fig:sm-quantum-examples}
\end{figure}

The main text gives an example of the quantum Mpemba effect in a symmetric single well. Here we give another example with a simpler potential, the one-dimensional infinite square well, together with examples for the other three potential classes.
Figure~\ref{fig:sm-quantum-examples} shows $a_{2}(\beta)$ for each potential.

We take
\begin{equation}
  V(x)=
  \begin{cases}
    0, & |x|<L/2,\\
    \infty, & |x|\geq L/2.
  \end{cases}
  \qquad
  E_n=E_*n^2,
  \qquad E_*=\frac{\pi^2\hbar^2}{2ML^2},
  \quad n=1,2,\ldots.
  \label{eq:sm-infinite-well}
\end{equation}
Unlike the harmonic oscillator, position coupling connects
nonadjacent levels of opposite parity. For $m\ne n$,
\begin{equation}
  |x_{mn}|=
  \begin{cases}
    \displaystyle\frac{8Lmn}{\pi^2(m^2-n^2)^2}, & m+n\text{ odd},\\[4pt]
    0, & m+n\text{ even}.
  \end{cases}
  \label{eq:sm-infinite-well-matrix-elements}
\end{equation}
In units $\hbar=L=E_*=1$, the mass is
$M=\pi^2/2$. We set $\beta_b=1$ and artificially design a bath spectrum
\begin{equation}
  J(\omega)=\eta\omega
    \left[1+3000\exp\left(-\frac{(\omega-15)^2}{2(0.6)^2}\right)\right],
  \label{eq:sm-infinite-well-spectrum}
\end{equation}
which enhances the allowed $4\to1$ transition, with energy difference $E_4-E_1=15$. As shown in Fig.~\ref{fig:sm-quantum-examples}(a), $a_2(\beta)$ shows a nonmonotonic behavior, which is a signature of the quantum Mpemba effect.

For Fig.~\ref{fig:sm-quantum-examples}(b)--(d), we use $\hbar=k_{\mathrm B}=1$, $\hat H=-\hbar^2\partial_x^2/(2M)+V(x)$, $\beta_b=10$, and $J(\omega)=\eta\omega^3\ee^{-(\omega/\omega_c)^2}$. 
The potentials are
\begin{equation}
  \begin{aligned}
    \text{(b)}\quad &V(x)=(1-\ee^{-3x})^2+0.001x^2,\\
    \text{(c)}\quad &V(x)=\frac{2(x^2-1)^2}{1+x^2/L^2},\\
    \text{(d)}\quad &V(x)=\frac{(x^2-1)^2}{1+x^2/L^2}-0.2x,
  \end{aligned}
  \label{eq:sm-example-potentials}
\end{equation}
where $L=3$ for both double-well potentials.
As one can see, $a_2(\beta)$ is nonmonotonic in all cases, which indicates the presence of the quantum Mpemba effect.

\subsection{Wigner function}
\label{sec:sm-wigner-examples}

\begin{figure}[tbp]
\centering
\includegraphics[width=0.8\linewidth]{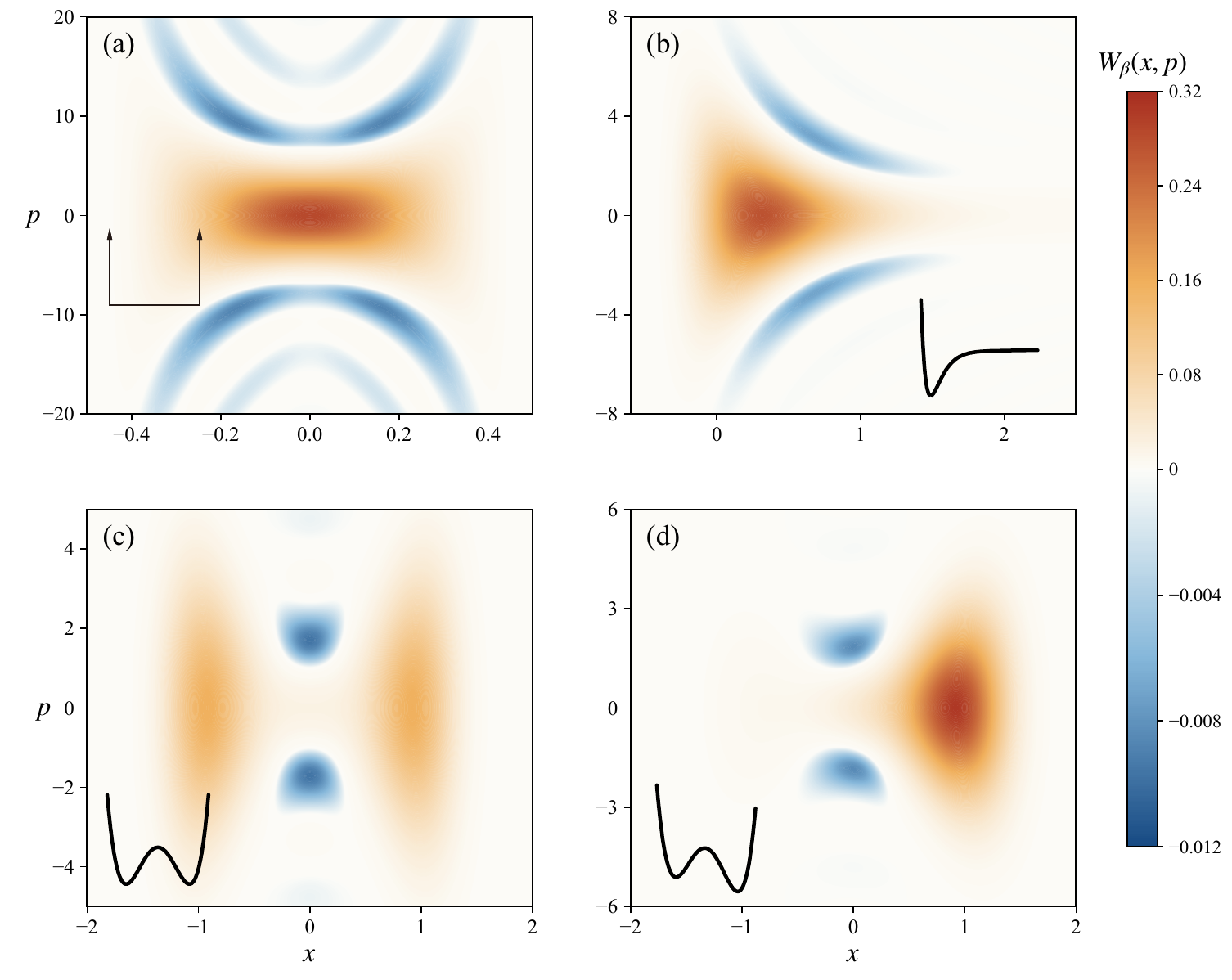}
\caption{The thermal Wigner function $W_\beta(x,p)$ for (a) infinite square well, (b) asymmetric single well, (c) symmetric double well, and (d) asymmetric double well.
The potentials are the same as in Fig.~\ref{fig:sm-quantum-examples}.
We set $\hbar=1$ in all panels, with $(M,\beta)=(\pi^2/2,1)$ in (a) and $(6,10)$ in (b)--(d).
The well width is $L=1$ in (a), and $L=3$ for the double-well potentials in (c) and (d).
Blue indicates negative values, while orange and red indicate positive values.
All panels share one color scale, with separate linear scales below and above zero.}
\label{fig:sm-wigner-examples}
\end{figure}

For each potential in Fig.~\ref{fig:sm-quantum-examples}, the thermal populations $\pi_\mu(\beta)$ define the density operator
$\rho_\beta=\sum_\mu\pi_\mu(\beta)|\mu\rangle\langle\mu|$.
Its Wigner function is
\begin{equation}
  W_\beta(x,p)=\frac{1}{2\pi\hbar}
  \int_{-\infty}^{\infty}\!\dd y\,\ee^{-\mathrm i py/\hbar}
  \left\langle x+\frac{y}{2}\right|\rho_\beta
  \left|x-\frac{y}{2}\right\rangle,
  \label{eq:sm-thermal-wigner}
\end{equation}
with $\int\!\dd x\,\dd p\,W_\beta(x,p)=1$.
We evaluate the thermal density matrix from the energy eigenstates,
\begin{equation}
  \langle x|\rho_\beta|x'\rangle
    =\sum_\mu\pi_\mu(\beta)\psi_\mu(x)\psi_\mu^*(x'),
  \qquad
  \pi_\mu(\beta)=\frac{\ee^{-\beta E_\mu}}{Z_q(\beta)},
  \label{eq:sm-thermal-density-kernel}
\end{equation}
and substitute it into Eq.~\eqref{eq:sm-thermal-wigner}.

Figure~\ref{fig:sm-wigner-examples} shows the resulting Wigner functions at $\beta=1$ for the infinite square well and $\beta=10$ for the other three potentials.
The masses are $M=\pi^2/2$ for the infinite square well and $M=6$ for the other three potentials.
All four thermal states have regions where $W_\beta(x,p)<0$.
These negative regions show that a state can remain diagonal in the energy basis while displaying Wigner negativity.
As established in Sec.~\ref{sec:sm-wigner-transitions}, this requires the position operator to connect nonadjacent energy levels.

\subsection{Classical asymmetric double well}
\label{sec:sm-classical-double-well-example}

\begin{figure}[tbp]
\centering
\includegraphics[width=0.8\linewidth]{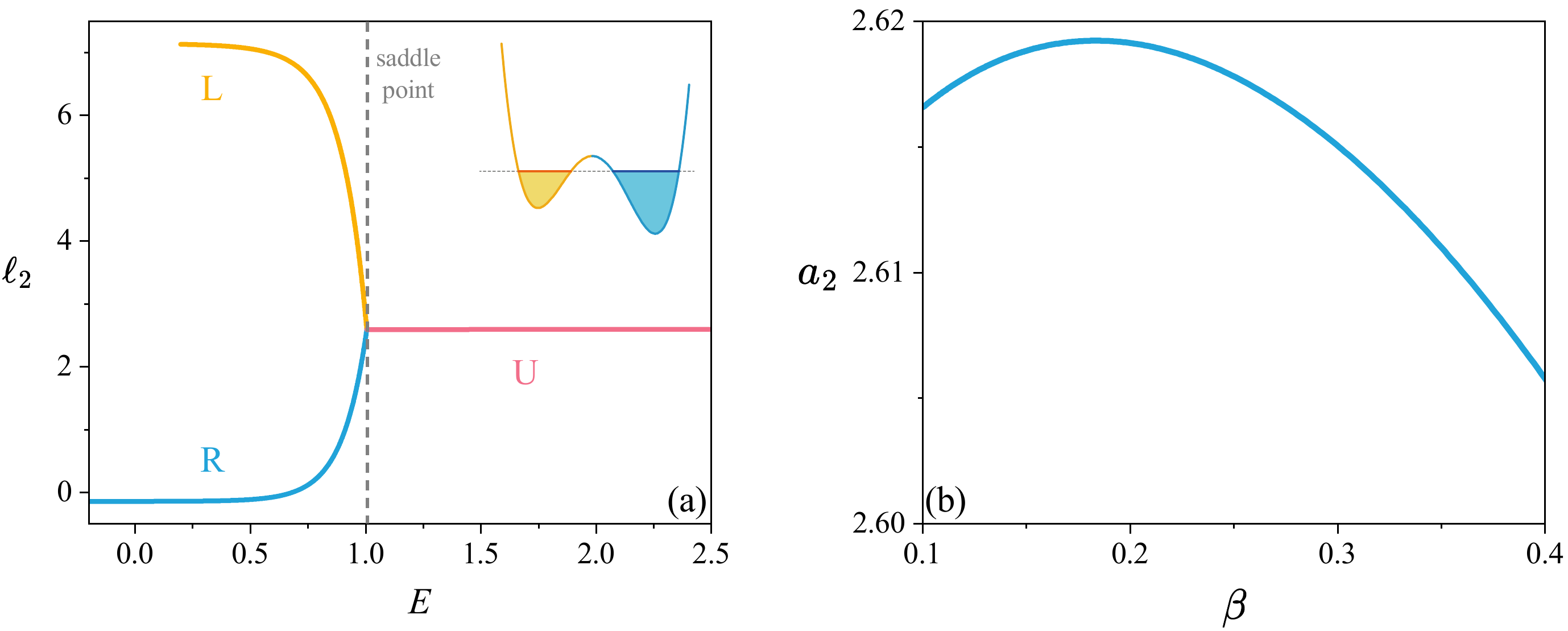}
\caption{The slowest left mode $\ell_{2,\alpha}(E)$ in (a) and the overlap coefficient $a_2(\beta)$ in (b) for the classical asymmetric double well. 
The branches $L$, $R$, and $U$ describe motion in the left well, right well, and above the barrier, respectively. 
The dashed line marks the barrier energy, and the inset sketches the potential without coordinate scales. 
We set $M=17$, $\beta_b=10$, $\eta=1$, $\omega_c=8$ and $L=3$.
} 
\label{fig:sm-classical-double-well}
\end{figure}

For the classical case, we use the same potential, bath spectrum and parameters as in Fig.~\ref{fig:sm-quantum-examples}(d).
Below the energy of the saddle point $E_s$, each periodic orbit of the isolated system lies within one well. We label the left- and right-well branches by $L$ and $R$.
Above $E_s$, the particle moves across both wells on the branch $U$.
For each branch $\alpha=L,R,U$, we define the density of states $g_\alpha(E)=2\pi/\Omega_\alpha(E)$, where $\Omega_\alpha(E)$ is the orbital angular frequency on that branch.
The thermal energy density on each branch is
\begin{equation}
  \pi_\alpha(E;\beta)=\frac{g_\alpha(E)\ee^{-\beta E}}{Z_c(\beta)},
  \qquad
  Z_c(\beta)=\sum_{\alpha=L,R,U}\int_\alpha\!\dd E\,g_\alpha(E)\ee^{-\beta E},
  \label{eq:sm-classical-branch-equilibrium}
\end{equation}
where each integral runs over the energy interval of its branch, so that $Z_c(\beta)$ normalizes the probability over all three branches.

We calculate the diffusion coefficient on each branch from its periodic orbits using Eq.~\eqref{eq:sm-classical-diffusion-coefficient}, and solve Eq.~\eqref{eq:sm-classical-energy-diffusion} separately on the three energy intervals.
At $E_s$, we match the solutions by requiring continuity of $P_\alpha(E,t)/\pi_\alpha(E;\beta_b)$ and conservation of probability current.
We impose zero probability current at the lowest energy of each well.
These conditions connect the three branch equations into one evolution operator, from which we obtain the slowest left mode $\ell_{2,\alpha}(E)$.
Then, we normalize it by $\sum_\alpha\int_\alpha\!\dd E\,\pi_\alpha(E;\beta_b)\ell_{2,\alpha}^2(E)=1$, with zero mean at equilibrium.

Figure~\ref{fig:sm-classical-double-well}(a) shows this mode on the three branches. 
The two well branches have different values at the same energy and meet the upper branch at $E_s$.
Thus $\ell_2$ depends on the branch as well as the energy, and the monotonicity constraint for a single energy interval does not apply.
For a thermal initial state, the overlap coefficient is
\begin{equation}
  a_2(\beta)=\sum_{\alpha=L,R,U}a_{2,\alpha}(\beta),
  \qquad
  a_{2,\alpha}(\beta)=\int_\alpha\!\dd E\,
    \ell_{2,\alpha}(E)\pi_\alpha(E;\beta),
  \label{eq:sm-classical-branch-overlap}
\end{equation}
where each integral runs over the energy interval of its branch.
As shown in Fig.~\ref{fig:sm-classical-double-well}(a), $\ell_2$ takes larger values on the $L$ branch than on the $U$ branch. 
As the temperature increases, thermal probability shifts from the two well branches to the upper branch. This redistribution may reduce the overlap coefficient $a_2(\beta)$ and lead to a classical Mpemba effect. 
As illustrated in Fig.~\ref{fig:sm-classical-double-well}(b), $a_2(\beta)$ is nonmonotonic in $\beta$, which indicates the presence of the classical Mpemba effect.

\subsection{Numerical convergence}
We check energy truncation by monitoring $\lambda_2$ and $a_2(\beta)$ as the retained basis is enlarged. For Fig. 2 of the main text, we retain $1400$ levels up to $E_{\max}=180$, corresponding to $\beta(E_{\max}-E_0)\simeq17.8$ at $\beta=0.1$. Increasing $E_{\max}$ from $140$ to $180$ changes $\lambda_2$ by less than $10^{-10}$ relatively and the independently peak-normalized $|a_2|$ curves by less than $2\times10^{-5}$ over $\beta\in[0.1,4]$. The KL dynamics uses $587$ levels up to $E_{\max}=80$. Increasing this cutoff from $60$ to $80$ changes the crossing time by less than $10^{-9}$ relative.

For Fig. 3, we retain $500$, $1200$, and $4900$ levels up to $E_{\max}=20$ for $\hbar=1$, $0.4$, and $0.1$, respectively. For these three curves, increasing $E_{\max}$ from $16$ to $20$ changes $\lambda_2$ by less than $10^{-10}$ relatively and the independently peak-normalized curves by at most $8.4\times10^{-3}$. For the critical curve at $\hbar_c\approx0.83$, we retain $800$ levels up to $E_{\max}=26$ with spatial grid spacing $\Delta x=0.0025$. At fixed $E_{\max}=26$, halving $\Delta x$ from $0.005$ to $0.0025$ shifts the estimated threshold from $\hbar_c=0.82851007$ to $0.82849402$, a change of approximately $1.6\times10^{-5}$.

For Fig.~\ref{fig:sm-quantum-examples}(a)-(d), we retain $30$, $500$, $100$, and $200$ levels, respectively. For the infinite square well, increasing the basis from $30$ to $100$ levels changes $\lambda_2$, the peak amplitude, and the selected hot-state amplitude by less than $2\times10^{-12}$ relative. For the two quantum double-well potentials, increasing the energy cutoff from $100$ to $160$ and halving the spatial grid spacing changes $\lambda_2$ by less than $3\times10^{-6}$ relatively and $a_2$ by less than $2\times10^{-5}$ of its reference peak value. The classical calculations use upper energy boundaries of $40$ for Fig. 3 and $160$ for Fig.~\ref{fig:sm-classical-double-well}. Comparisons with smaller energy domains and coarser grids give maximum changes in $a_2$ below $10^{-3}$ of the respective reference peak values.